\documentclass[11pt,twocolumn,aip,apl,reprint]{revtex4-1}

\newcommand{\addDB}[1]{\textcolor{black}{#1}}
\newcommand{\addDS}[1]{\textcolor{black}{#1}}

\usepackage{graphicx,amsmath,amssymb,color}

\begin{document}

\title{Dual Origin of Room Temperature Sub-Terahertz Photoresponse in Graphene Field Effect Transistors}

\author{D.A. Bandurin}
\affiliation{School of Physics and Astronomy, University of Manchester, Manchester, M13 9PL, UK}

\author{I. Gayduchencko}
\affiliation{Physics Department, Moscow State University of Education (MSPU), Moscow, 119435, Russian Federation}

\author{Y. Cao}
\affiliation{School of Physics and Astronomy, University of Manchester, Manchester, M13 9PL, UK}

\author{M. Moskotin}
\affiliation{National Research University of Electronic Technology, 1, Shokin Square, Zelenograd, Moscow, 124498, Russia}

\author{A. Principi}
\affiliation{School of Physics and Astronomy, University of Manchester, Manchester, M13 9PL, UK}

\author{I.V. Grigorieva}
\affiliation{School of Physics and Astronomy, University of Manchester, Manchester, M13 9PL, UK}

\author{G. Goltsman}
\affiliation{Physics Department, Moscow State University of Education (MSPU),Moscow, 119435, Russian Federation}

\author{G. Fedorov}
\affiliation{Moscow Institute of Physics and Technology (State University), Dolgoprudny 141700, Russia}

\author{D.~Svintsov}
\affiliation{Moscow Institute of Physics and Technology (State University), Dolgoprudny 141700, Russia}
\email{denis.bandurin@manchester.ac.uk}

\begin{abstract}
Graphene is considered as a promising platform for detectors of high-frequency radiation up to the terahertz (THz) range due to graphene$'$s superior electron mobility. Previously it has been shown that graphene field effect transistors (FETs) exhibit room temperature broadband photoresponse to incoming THz radiation thanks to the thermoelectric and/or plasma wave rectification. Both effects exhibit similar functional dependences on the gate voltage and therefore it was found to be difficult to disentangle these contributions in the previous studies. In this letter, we report on combined experimental and theoretical studies of sub-THz response in graphene field-effect transistors analyzed at different temperatures. This temperature-dependent study allowed us to reveal the role of photo-thermoelectric effect, p-n junction rectification, and plasmonic rectification in the sub-THz photoresponse of graphene FETs. 
\end{abstract}

\maketitle

Over the last decade, graphene has attracted a considerable attention in the fields of photonics\cite{bonaccorso2010photonics}, plasmonics\cite{grigorenko2012plasmonics}, and optoelectronics\cite{koppens2014photodetectors}. The interest is motivated by graphene’s unique gate-tuneable physical properties that allow realization of radiation detectors operating in a wide range of frequencies~\cite{Gabor2011HotCarrier,Jung2016Microwave,Fuhrer2012bolometer,cai2014sensitive}.

Electromagnetic radiation in the terahertz (THz) range deserves a special attention as it allows fast and non-destructive imaging of objects with a strong potential in medical and security sectors\cite{mittleman2013sensing}. With this potential, the development of efficient THz generators and sensitive detectors is an important technological problem. 

Recently, it has been shown that graphene field-effect transistors (FETs) can act as THz detectors exhibiting a dc photoresponse to impinging radiation~\cite{cai2014sensitive,vicarelli2012graphene,Spirito2014bilayer,Tong2015AntennaEnhanced,auton2016rectifier,Auton2017RectifiersImaging,generalov2017400,zak2014antenna,qin2017room}. A broadband photodetection in the sub-THz range with the responsivity reaching tens of V/W and noise equivalent power of hundreds of pW/Hz$^{1/2}$ has been demonstrated in graphene FETs designed in the configuration where the incoming radiation is coupled between the source and the gate terminals~\cite{vicarelli2012graphene,Spirito2014bilayer,generalov2017400,zak2014antenna}. In this configuration, the photoresponse is usually attributed to the so-called Dyakonov-Shur (DS) rectification arising as a result of the plasma waves excitation in the FET channel~\cite{dyakonov1996detection,Tomadin2013Plasma}. However, other effects can also impact the photoresponse. For instance, photo-thermoelectric effect (PTE) arising from the temperature gradient in a FET can provide an additional rectification of the incoming high-frequency signal~\cite{Jung2016Microwave,cai2014sensitive}.
\addDB{As we show below, both PTE and DS effects exhibit similar functional dependence on the gate voltage and result in the same sign of the photoresponse, that makes it  challenging to point to the origin of the observed rectification. Further improvement of graphene-based THz photodetectors requires a deeper understanding of the rectification mechanisms governing the photoresponse. }

%Most of the previous works were performed at room temperature (RT) and analyzed the photoresponse dependence on the gate voltage only \addDB{that, as we show below, is insufficient to argue about the origin of the photoresponse. In addition, the sign of the photovoltage was argued to be opposite for the PTE and DS rectification which was considered a limiting factor for the performance of graphene photodetectors. } For further improvement of graphene-based THz detectors a deeper understanding of the mechanisms governing the photoresponse is required. 

% \addDB{In addition, it was discussed that the PTE and DS produce the photovoltage of opposite polarities limiting the performance of graphene-based photodetectors}.

\addDB{In this work, we analyze the sub-THz photoresponse of graphene-based FET by comparing its responsivity at liquid nitrogen and room temperatures. Such temperature-dependent measurements allowed us to point to the effects arising from the heating of electronic system, overdamped plasma wave photodetection, and diode rectification. In particular, we show that the opposite sign of the Seebeck coefficients in the p-doped graphene channel and n-doped graphene in the vicinity of the metal contacts causes a significant PTE photoresponse to incoming radiation. When graphene channel is uniformly n-doped we observe an enhancement of the photoresponsivity with increasing temperature that can be explained by the overdamped DS scenario. } 

Our FET was made of graphene encapsulated between two slabs ($50$ nm thick each) of hexagonal boron nitride (hBN) by the dry-transfer method as described in Supporting Information and elsewhere~\cite{kretinin2014electronic,Wang2013onedimensional}. The FET was made in a dual-gated configuration such that the carrier density $n$ in the channel was controlled by the global back gate electrode (located at a distance $d=500$ nm) whereas the top-gate as well as the source terminals were extended to a millimeter scale and served as sleeves of a logarithmic spiral broadband antenna, see Fig. 1a-b and Supporting Information, section I.

Photoresponse measurements were performed in a variable temperature optical cryostat allowing coupling of the device under study to electromagnetic radiation via a polyethylene window. A silicon hemispherical lens focusing the incoming radiation to the device antenna was attached to the bottom of the device. The radiation was further funneled into the channel of our FET by its coupling to the source and the top gate electrodes yielding the modulation of the top gate-to-channel voltage difference, Fig. 1a. The sub-THz radiation was generated by two backward wave oscillators allowing us to tune the frequency between $f=0.13$ THz and $0.45$ GHz.  The power delivered to the device was measured using the Golay cell and recalculated accounting for the losses and size of the cryostat optical window and silicon lens\cite{fedorov2013photothermoelectric}.

\begin{figure}[ht!]
\centering\includegraphics[width=1.0\linewidth]{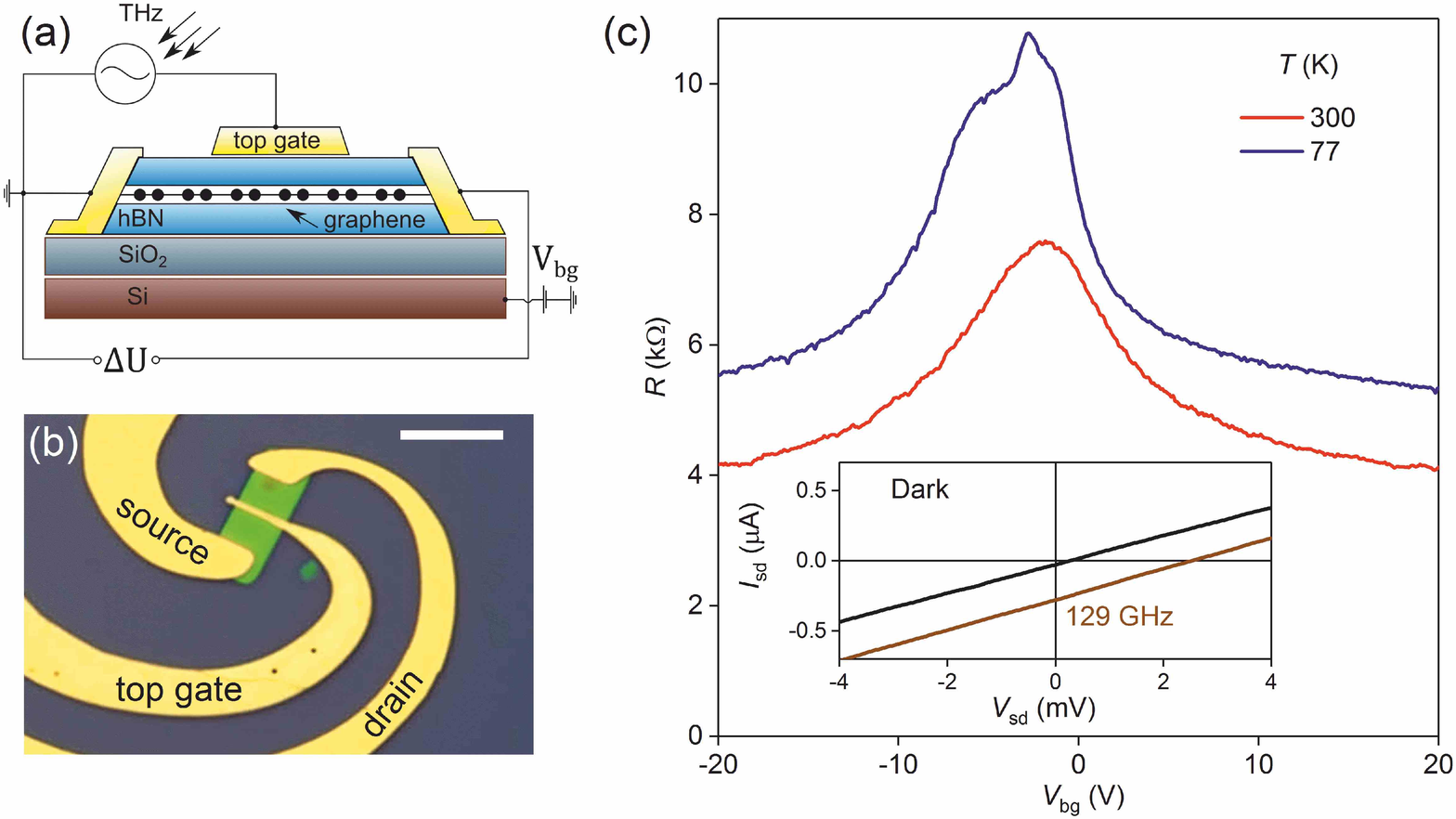}
\caption{ (a) Schematics of a dual-gated graphene-based field effect transistor (b) An optical micrograph of our device. Scale bar is 6 $\mu$m (c) Two-terminal resistance as a function of $V_{bg}$ measured at $T=77$ K and $300$ K. Inset: examples of the $I(V)-$ curves measured in the dark and under illumination with $0.13$ THz. $V_{bg}=-20$ V,  $T=300$ K.}
\end{figure}

Prior to photoresponse measurements, we characterized the transport properties of our graphene FET. Figure 1c shows a two-terminal resistance $R$ as a function of back gate voltage $V_{bg}$ measured at $T=77$ and $300$ K. At RT, $R$ exhibits a peak of 7 k$\Omega$ located around $V_{bg}=-2$ V corresponding to the charge neutrality point (CNP). Away from the CNP, $R$ remains above several k$\Omega$. At lower $T$, we observed an increase of the device resistance. Such behavior is opposite to that expected for doped graphene. This is not surprising as all our measurements were carried out in a two-terminal geometry and, therefore, $R$ also accounts for a temperature-dependent contact resistance. For the same reason, the RT field effect mobility $\mu$, extracted from the slope of $R(V_{bg})$ dependence, of our graphene device was only of $3.2\times 10^3$ cm$^2$V$^{-1}$s$^{-1}$: two-terminal measurements provide the lower bound for $\mu$, the latter is usually higher in encapsulated samples~\cite{Wang2013onedimensional}. Nevertheless, at liquid nitrogen temperature we observed a two-fold increase of $\mu$ indicating the suppression of electron-phonon scattering leading to the increase of the scattering time $\tau$.

The inset to Fig. 1c provides examples of the RT $I(V)$-curves of our photodetector measured in the absence (black) and in the presence of incoming radiation (brown). Both curves demonstrate linear dependence of the source-drain current $I_{SD}$ on the bias voltage $V_{SD}$. The $I(V)$-curve measured under illumination is notably shifted to the right from the origin so that it intercepts the zero-current level at $\Delta U\approx 2.8$ mV, which we further refer to as the photovoltage.
\begin{figure}[ht!]
\centering\includegraphics[width=0.8\linewidth]{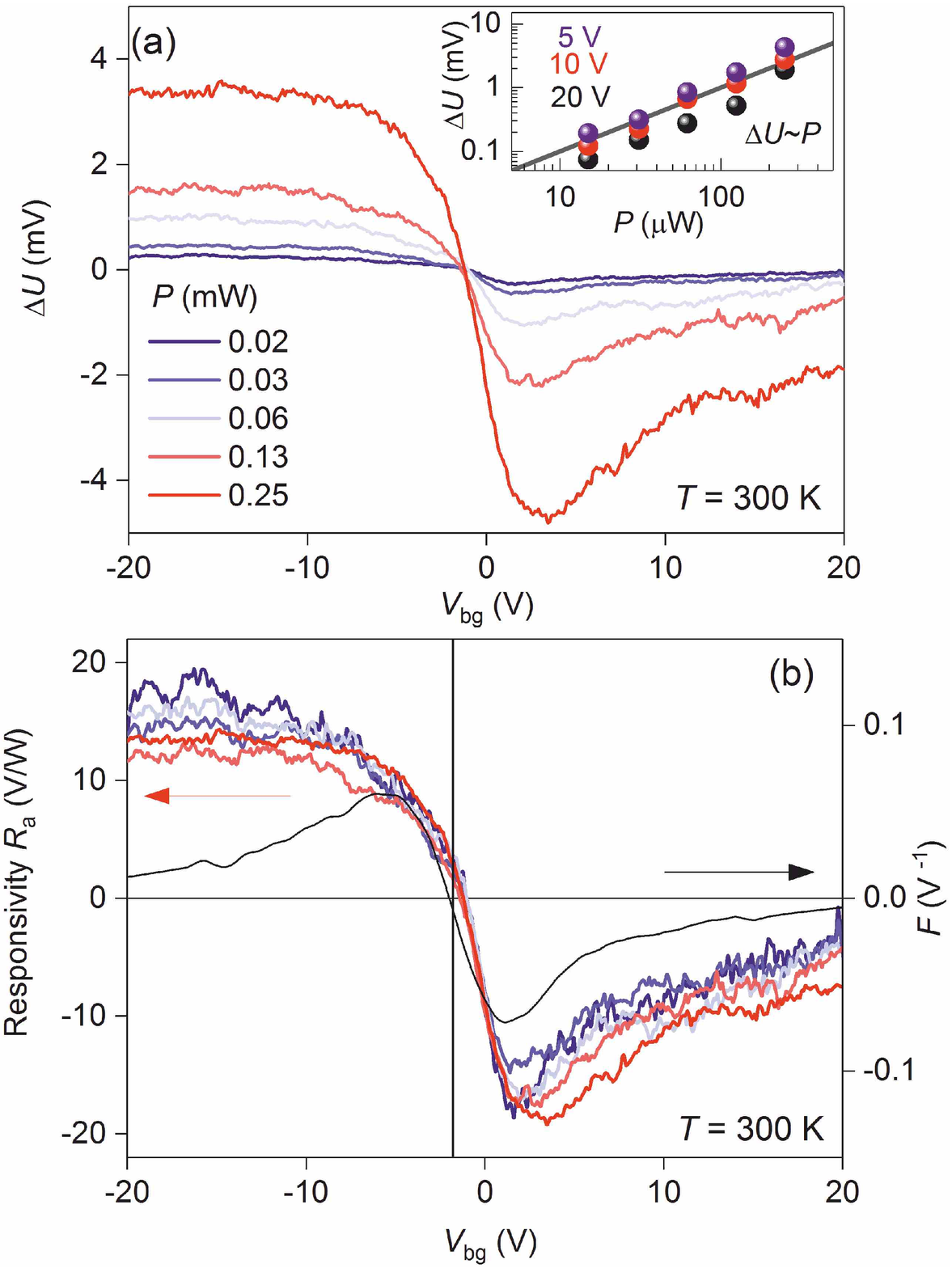}
\caption{ (a) Photovoltage as a function of $V_{bg}$ measured at varying power of incoming $0.13$ THz radiation. $T=300$ K. Inset: Photoresponse as a function of incoming power for different $V_{bg}$. The solid grey line represents $\Delta U \propto P$ scaling. (b) RT responsivity versus back gate voltage for different $P$. Black: Gate dependence of room-temperature FET-factor $F={-{\sigma }^{-1}}d\sigma /d{{V}_{bg}}$, where $\sigma $ is the channel conductance.}
\end{figure}

Figure 2a shows the results of the RT measurements of $\Delta U$ as a function of $V_{bg}$ acquired at different power $P$ of incoming 0.13 THz radiation. A finite $\Delta U$ is observed at all experimentally accessible $V_{bg}$ except CNP where the sign of the photoresponse changes in agreement with ambipolar transport in graphene. The detected signal is highly asymmetric with respect to $V_{bg}$, such that it tends to zero at large positive $V_{bg}$, whereas it remains nearly constant with decreasing $V_{bg}$ below CNP. 

\addDB{As the power of impinging radiation changes, the photoresponse voltage scales linearly with $P$, as shown in Fig. 2a. In Fig. 2b we plot the photoresponsivity $R_a=\Delta U/P$ as a function of $V_{bg}$ corresponding to $\Delta U$ shown in Fig. 2a. All the dependences fall onto the same line and do not depend on the power of incoming radiation indicating that the device remains in the linear-response regime for the experimentally accessible power-range. The maximum value of $R_a$ and the minimum noise equivalent power, NEP, of our device were 20 V/W and 0.6 nW/Hz$^{1/2}$ (see Supporting Information, section II) respectively that is comparable to the performance  reported previously\cite{vicarelli2012graphene,Spirito2014bilayer,generalov2017400,zak2014antenna}. For further characterization of our photodetector, we have also measured its responsivity at higher frequencies and found that our device exhibits a broadband photoresponse at all experimentally accessible gate voltages away from the CNP. The results of such measurements are presented in Supporting Information, section II. }

\begin{figure}[ht!]
\centering\includegraphics[width=0.8\linewidth]{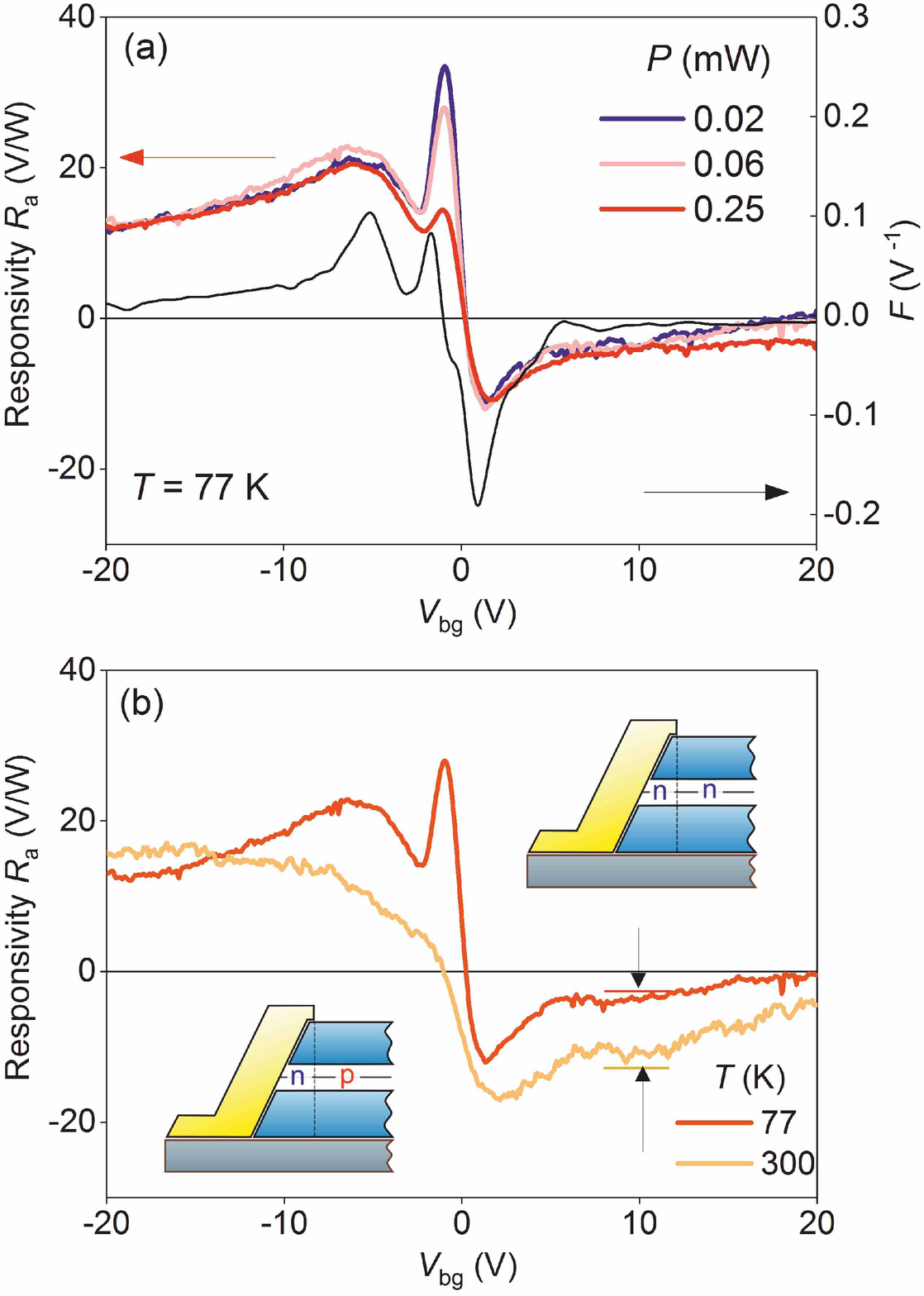}
\caption{ (a) Responsivity as a function of $V_{bg}$ measured at $T=77$ K for different power of incoming 0.13 THz radiation (b) $R_a (V_{bg})$ at different temperatures measured at $P=0.06$ mW. Arrows indicate the difference in $R_a$ between the data acquired at $T=77$ and $300$ K}
\end{figure}

Figure 3a shows the responsivity as a function of $V_{bg}$ measured at $T=77$ K for varying power of incoming radiation. Similarly to the RT measurements, $R_a$ is highly asymmetric with respect to positive and negative gate voltages. The photoresponse measured away from the CNP on a hole-doping side exhibits power-independent $R_a$ slowly varying with $V_{bg}$. For n-doping, $R_a$ rapidly tends to zero with increasing $V_{bg}$. Interestingly, at positive $V_{bg}$, the responsivity measured at $T=77$ K was found to be smaller than that obtained at RT, Fig. 3b.

The gate dependence of $R_a$ at $77$ K peaks slightly to the left from the CNP, and for the lowest power of radiation reaches $30$ V/W. In this gate voltage range, the $R(V_{bg})$ dependence can be represented as a superposition of two bell-shaped curves. \addDB{Usually, such a double-peak structure of the $R(V_{bg})$ dependence is caused by the $p$-$n$ junction formed at the boundaries between the regions with slightly different doping. In our device, those regions correspond to the areas not covered by the top-gate and the part of graphene under the top-gate in which a build-in electric field produces an unintentional doping.} Thus, the spike in $R_a$ can be attributed to the rectification of the high-frequency signal by this $p$-$n$ junction~\cite{vasilyev2017high}. The peak disappears with increasing $P$ which is in agreement with $p$-$n$ junction rectification scenario, as hot electrons pass over the junction barrier freely which suppresses rectification.

Further improvement of the THz detection efficiency requires a deeper understanding of the possible photoresponse mechanisms in graphene FETs which we discuss in detail below. Since graphene is characterized by the large optical phonon energy, hot electrons created by the Joule heating can remain at a higher temperature than that of the lattice and therefore may affect the photoresponse by the PTE~\cite{Gabor2011HotCarrier,Jung2016Microwave}. \addDS{Due to the asymmetric design of the antenna (the drain sleeve is much thinner than source and gate sleeves and is closely located to the latter), the high-frequency current flows predominantly between the source and the gate terminals (Supporting Information, section IV)}. This leads to the asymmetric temperature distribution across the graphene channel. Therefore, the electron system in the vicinity of the source remains at a higher average temperature $T_S$ compared to that near the drain $T_D$, as shown in Fig. 4a. For a uniformly doped channel, the emerging photovoltage is given by
\begin{equation}
\Delta U_{\rm PTE} = - \int{S dT} \approx S (T_S - T_D)
\end{equation}
where $S \propto T \sigma^{-1} d\sigma/dE_F$ is the Seebeck coefficient, $\sigma$ is the conductivity of graphene and $d\sigma/dE_F$ is its derivative with respect to the Fermi energy $E_F$. Within a factor of a slowly varying function $dV_{bg}/dE_F \propto V^{1/2}_{bg}$, the Seebeck coefficient and therefore $\Delta U_{\rm PTE}$ depend on the ratio between the 2D channel transconductance and its conductivity which we further refer to as the FET factor $F=-\sigma^{-1}  d\sigma/dV_{bg}$. In Figs. 2b and 3a we plot $F(V_{bg})$ obtained by numerical differentiation of the data in Fig. 1c. The experimental photoresponse follows $F(V_{bg})$ for positive gate voltages however strongly deviates from it when $V_{bg}<0$.

This asymmetry can be understood within the PTE model by taking into account $n$-doping of graphene near the gold contacts~\cite{nouchi2011determination,mccreary2010effect}. Considering the piecewise distribution of the Seebeck coefficient, $S_{cont}$ near the contacts and $S_{ch}$ in the remaining part of the channel, as shown in Fig. 4a, one obtains a modified expression for $\Delta U_{\rm PTE}$:
\begin{equation}
\label{PTE-modified}
\Delta U_{\rm PTE}  = (S_{ch} - S_{cont})(T_S - T_D).
\end{equation}
Equation (\ref{PTE-modified}) readily explains the asymmetry of the detector characteristics $R_{a}(V_{bg})$ shown in Fig. 2b and 3a. Indeed, $S_{ch}$ is controlled by the gate voltage and changes sign at the CNP while $S_{cont}$ is dictated only by the built-in field at the graphene-metal interface. The n-doping of graphene by contacts~\cite{nouchi2011determination,mccreary2010effect} implies that $S_{cont}<0$. This increases the absolute value of the photovoltage for $p$-doped channel and decreases it for $n$-doped . At high positive voltage $V_{bg} \ge 20$ V, the observed response approaches zero, which hints on the equal doping of channel and contacts. At this point, the electron Fermi energy in the channel is estimated as $E_F \approx \hbar v_F (\pi C_{bg} V_{vg})^{1/2} \approx 0.1$ eV (\addDB{where $v_F$ is the Fermi velocity and $C_{bg}$ is the gate-to-channel capacitance}),  which is close to the value of contact doping at graphene/metal interface~\cite{lee2008contact}.

An alternative view on the photodetection processes in 2D FETs is the so-called DS scenario which predicts the rectification of the ac signal by the field effect and hydrodynamic nonlinearities possibly enhanced by the resonant plasma wave excitations~\cite{dyakonov1996detection}. Though the conditions of the hydrodynamic transport are usually fulfilled in encapsulated graphene~\cite{Bandurin-HDgraphene}, the plasmonic enhancement of the nonlinearities in our device looks hardly possible as the plasma waves are overdamped. Indeed, for our device, we estimate the scattering time with respect to momentum non-conserving collisions $\tau$ to hundreds of fs so that $\omega \tau < 1$, where $\omega = 2\pi f$. In this non-resonant regime, the DS photodetection yields a broadband photoresponse \addDB{(also referred to as resistive self-mixing)} with the photovoltage given by~\cite{dyakonov1996detection,vicarelli2012graphene}:
\begin{equation}
\Delta U_{DS} = - \frac{U_a^2}{4}\frac{1}{\sigma}\frac{d\sigma}{d V_{g}} g(\omega) \propto F(V_{g})
\end{equation}
where $U_a$ is the ac gate-to-source voltage, $L$ is the channel length, $g(\omega) = [\sinh^2kL -\sin^2 kL]/[\sinh^2kL +\cos^2 kL]$ is the “form factor” depending on the wave number $k = (\omega/2\tau)^{1/2}/s$ of the overdamped plasma wave, $s = \sqrt{e|V_{bg}|/m}$ is the plasma wave velocity, \addDB{$e$ is the electron charge,} and $m$ is the \addDB{cyclotron mass of charge carriers}. Importantly, $\Delta U_{DS}$ is also proportional to $F$, that makes it difficult to distinguish the PTE and DS mechanisms by studying the gate-voltage dependences only, see equations (1) and (3). \addDB{We also emphasize, that these two mechanisms yield the same sign of the photovoltage (see Supporting Information). As we show below, the role of each effect in the photovoltage can be revealed by studying the photoresponse at different temperatures. }

\addDB{We stat our analysis from the temperature dependence of the PTE. To this end, we express the electron temperature in eq. (\ref{PTE-modified}) through the incoming radiation power by solving the heat transfer equation (Supplementary Material, section III).} This results in the responsivity
\begin{equation}
R_a \approx \frac{3}{2\pi^2}\left[\frac{e}{k_B}(S_{cont} - S_{ch})\right] \frac{e|Z_{\rm eff}|}{k_B T}\frac{\delta L}{L},
 \end{equation}
where $Z_{\rm eff}$ is the effective resistance \addDB{relating the power impinging on the antenna with resulted gate-to-channel voltage}, $\delta L$ is the length of doped contact region and $L$ is the overall channel length. In Fig. 4b we show the normalized responsivity $R_{a}/|Z_{\rm  eff}|$ as calculated from eq. (4) for different temperatures and realistic estimates of $\delta L=100$ nm~\cite{allen2015visualization,shalom2016quantum}. At negative $V_{bg}$, both curves saturate and exhibit temperature-independent response. When approaching the CNP from the p-doping side, one observes an increase in responsivity with lowering temperature. Both observations are in accordance with the experimental data for p-doped graphene shown in Fig. 3b. The model allows us to roughly estimate  $|Z_{\rm  eff}|\approx 100$ $\Omega$, which is close to that estimated from the equivalent circuit (see Supporting Information). 

% as it follows from equivalent circuit of antenna-coupled sample (supporting information, section IV).\addDB{Do we need this sentence and this part of supplementary? Shall we just remove the real estimate and leave it qualitatevely?}

\begin{widetext}

\begin{figure}[t]
\centering\includegraphics[width=0.8\linewidth]{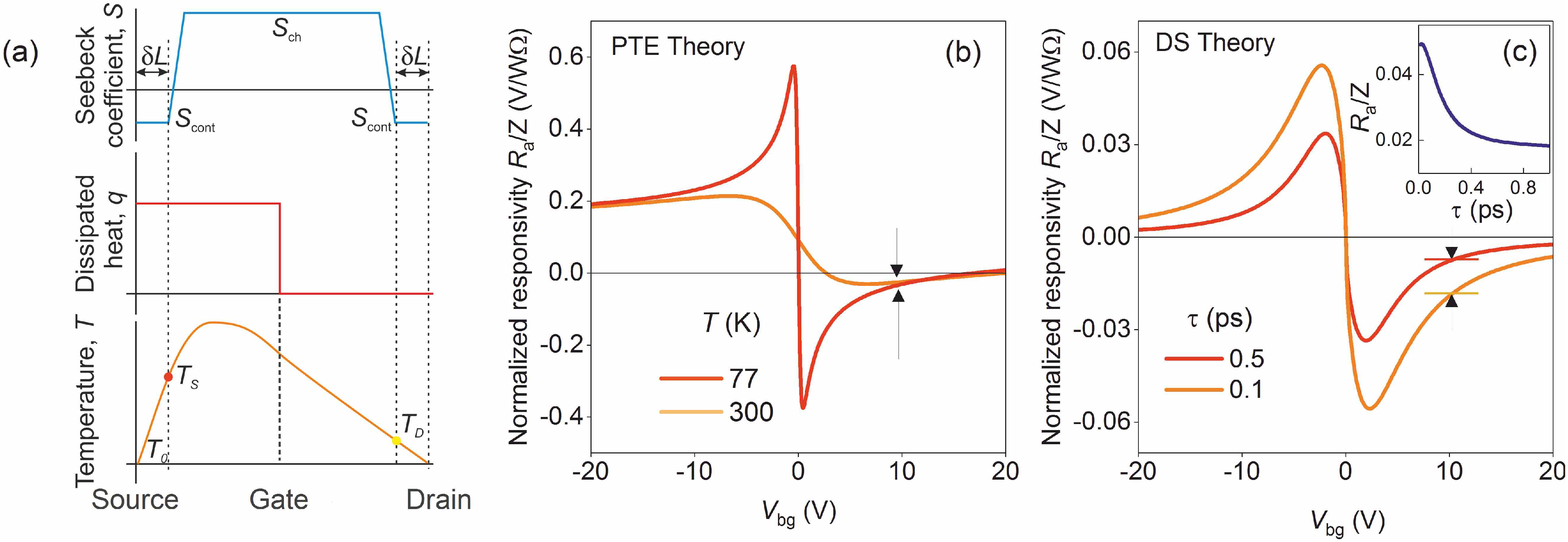}
 \caption{ (a) Distribution of the Seebeck coefficient $S$, dissipated heat $q$ and temperature $T$ in a FET with the radiation coupled between the source and gate terminals. (b) Normalized photo-thermoelectric responsivity as a function of $V_{bg}$ calculated for $T=77$ and $300$ K.
(c) Normalized DS photoresponsivity as a function of $V_{bg}$  calculated with eq. (3) for given $\tau$ and $f=0.13$ THz. Inset: $R_a/Z$ as a function of $\tau$ calculated  at $V_{bg}=-5$ V. \addDB{ $V_{bg}=0$ V corresponds to the CNP. }}
\end{figure}
\end{widetext}

Despite the agreement with data at negative gate voltages, the PTE theory does not provide a complete qualitative description of the experiment: in the $n$-doped regime, the absolute value of the RT responsivity is three times higher than that observed at $T=77$ K which contradicts the PTE scenario, because the latter yields the decrease of the photovoltage with increasing temperature, Fig. 4b~\cite{xu2009photo}. Indeed, in the case of PTE, the relative heating $T_S - T_D$ is proportional to the dissipated heat $q=\sigma E^2/2$ and inversely proportional to the thermal conductance $\chi$. At liquid nitrogen temperatures and away from CNP, the electric and thermal conductivity are coupled by the Wiedemann-Franz law, and therefore $T_S - T_D \propto q/\chi \propto T^{-1}$. Recalling that in the degenerate electron system the Seebeck coefficient grows linearly with $T$, one obtains the temperature independent $\Delta U_{\rm PTE}$. In the non-degenerate case $S$ weakly depends on temperature while the relative heating and therefore $\Delta U_{\rm PTE}$ drop upon increasing $T$. Further reduction of the photovoltage at elevated temperatures occurs due to electron-phonon cooling~\cite{crossno2015thermometry,Fong2013} and interaction with hyperbolic phonon-polaritons in hBN~\cite{tielrooij2017out}.

The increase of the responsivity in $n$-doped graphene at RT can be qualitatively explained by the broadband DS photodetection scenario. Indeed, although the eq. (3) does not depend on temperature explicitly, the latter enters the expression for the plasma wave vector $k$ through the temperature-dependent scattering rate $\tau$~\cite{dyakonov1996detection}. The evaluation of the DS response with eq. (3) for $f=0.13$ THz including the form factor $g(\omega)$ for different relaxation times demonstrates a counterintuitive increase in signal at shorter $\tau$, as shown in Fig. 4c, that is in qualitative agreement with Fig. 3b. Previously we mentioned that in our device the momentum relaxation time drops with increasing temperature (e.g. due to electron-phonon scattering). Moreover, the PTE response is expected to vanish for highly n-doped device (Fig. 4b). Therefore, an increase of the experimentally detected responsivity at RT may be caused by the overdamped DS photoresponse being the dominant photodetection mechanism. We also note that for an accurate evaluation of the DS photovoltage in such FET, one needs to account for its dual-gated design, that is the subject of our further studies.

In summary, we have demonstrated that FETs based on graphene encapsulated in hexagonal boron nitride can serve as high-responsivity sub-THz photodetectors with low NEP. \addDB{We have shown, how the measurements of the photoresponse at different temperatures can provide the information on the rectification mechanisms governing the photoresponse.} Namely, by comparing the photoresponse at different temperatures, we have found that the opposite sign of the Seebeck coefficients in the p-doped graphene channel and n-doped graphene-metal interface results in the significant PTE rectification of the high-frequency radiation. For a uniformly n-doped graphene (where the PTE is \addDB{minimized}), we have found that the photoresponse increases with increasing temperature that contradicts the PTE scenario and can be qualitatively explained by the overdamped plasma wave rectification being the dominant contribution to the photovoltage at RT. \addDB{Furthemore, we have found that a $p$-$n$ junction, formed in the channel of a FET, provides an additional rectification mechanism allowing one to increase the responsivity of graphene-based THz photodetectors. }
\newline

D.A.B. acknowledges Leverhulme Trust for financial support. The work of DS was supported by the grant \#16-19-10557 of the Russian Scientific Foundation. G.F.,I.G., M.M. and G.G acknowledge Russian Science Foundation (Grant No.14-19-01308, MIET and Grant No. 17-72-30036, MSPU), Ministry of Education and Science of the Russian Federation (contract No. 14.B25.31.0007 and Task No. 3.7328.2017/LS) and Russian Foundation for Basic Research (grant no. 15-02-07841). The authors are grateful to prof. M.S. Shur for helpful discussions.

\bibliography{Bibliography}

\begin{thebibliography}{34}%
\makeatletter
\providecommand \@ifxundefined [1]{%
 \@ifx{#1\undefined}
}%
\providecommand \@ifnum [1]{%
 \ifnum #1\expandafter \@firstoftwo
 \else \expandafter \@secondoftwo
 \fi
}%
\providecommand \@ifx [1]{%
 \ifx #1\expandafter \@firstoftwo
 \else \expandafter \@secondoftwo
 \fi
}%
\providecommand \natexlab [1]{#1}%
\providecommand \enquote  [1]{``#1''}%
\providecommand \bibnamefont  [1]{#1}%
\providecommand \bibfnamefont [1]{#1}%
\providecommand \citenamefont [1]{#1}%
\providecommand \href@noop [0]{\@secondoftwo}%
\providecommand \href [0]{\begingroup \@sanitize@url \@href}%
\providecommand \@href[1]{\@@startlink{#1}\@@href}%
\providecommand \@@href[1]{\endgroup#1\@@endlink}%
\providecommand \@sanitize@url [0]{\catcode `\\12\catcode `\$12\catcode
  `\&12\catcode `\#12\catcode `\^12\catcode `\_12\catcode `\%12\relax}%
\providecommand \@@startlink[1]{}%
\providecommand \@@endlink[0]{}%
\providecommand \url  [0]{\begingroup\@sanitize@url \@url }%
\providecommand \@url [1]{\endgroup\@href {#1}{\urlprefix }}%
\providecommand \urlprefix  [0]{URL }%
\providecommand \Eprint [0]{\href }%
\providecommand \doibase [0]{http://dx.doi.org/}%
\providecommand \selectlanguage [0]{\@gobble}%
\providecommand \bibinfo  [0]{\@secondoftwo}%
\providecommand \bibfield  [0]{\@secondoftwo}%
\providecommand \translation [1]{[#1]}%
\providecommand \BibitemOpen [0]{}%
\providecommand \bibitemStop [0]{}%
\providecommand \bibitemNoStop [0]{.\EOS\space}%
\providecommand \EOS [0]{\spacefactor3000\relax}%
\providecommand \BibitemShut  [1]{\csname bibitem#1\endcsname}%
\let\auto@bib@innerbib\@empty
%</preamble>
\bibitem [{\citenamefont {Bonaccorso}\ \emph {et~al.}(2010)\citenamefont
  {Bonaccorso}, \citenamefont {Sun}, \citenamefont {Hasan},\ and\ \citenamefont
  {Ferrari}}]{bonaccorso2010photonics}%
  \BibitemOpen
  \bibfield  {author} {\bibinfo {author} {\bibfnamefont {F.}~\bibnamefont
  {Bonaccorso}}, \bibinfo {author} {\bibfnamefont {Z.}~\bibnamefont {Sun}},
  \bibinfo {author} {\bibfnamefont {T.}~\bibnamefont {Hasan}}, \ and\ \bibinfo
  {author} {\bibfnamefont {A.}~\bibnamefont {Ferrari}},\ }\href@noop {}
  {\bibfield  {journal} {\bibinfo  {journal} {Nat. Photon.}\ }\textbf {\bibinfo
  {volume} {4}},\ \bibinfo {pages} {611} (\bibinfo {year} {2010})}\BibitemShut
  {NoStop}%
\bibitem [{\citenamefont {Grigorenko}, \citenamefont {Polini},\ and\
  \citenamefont {Novoselov}(2012)}]{grigorenko2012plasmonics}%
  \BibitemOpen
  \bibfield  {author} {\bibinfo {author} {\bibfnamefont {A.}~\bibnamefont
  {Grigorenko}}, \bibinfo {author} {\bibfnamefont {M.}~\bibnamefont {Polini}},
  \ and\ \bibinfo {author} {\bibfnamefont {K.}~\bibnamefont {Novoselov}},\
  }\href@noop {} {\bibfield  {journal} {\bibinfo  {journal} {Nat. Photon.}\
  }\textbf {\bibinfo {volume} {6}},\ \bibinfo {pages} {749} (\bibinfo {year}
  {2012})}\BibitemShut {NoStop}%
\bibitem [{\citenamefont {Koppens}\ \emph {et~al.}(2014)\citenamefont
  {Koppens}, \citenamefont {Mueller}, \citenamefont {Avouris}, \citenamefont
  {Ferrari}, \citenamefont {Vitiello},\ and\ \citenamefont
  {Polini}}]{koppens2014photodetectors}%
  \BibitemOpen
  \bibfield  {author} {\bibinfo {author} {\bibfnamefont {F.}~\bibnamefont
  {Koppens}}, \bibinfo {author} {\bibfnamefont {T.}~\bibnamefont {Mueller}},
  \bibinfo {author} {\bibfnamefont {P.}~\bibnamefont {Avouris}}, \bibinfo
  {author} {\bibfnamefont {A.}~\bibnamefont {Ferrari}}, \bibinfo {author}
  {\bibfnamefont {M.}~\bibnamefont {Vitiello}}, \ and\ \bibinfo {author}
  {\bibfnamefont {M.}~\bibnamefont {Polini}},\ }\href@noop {} {\bibfield
  {journal} {\bibinfo  {journal} {Nat. Nanotechnol.}\ }\textbf {\bibinfo
  {volume} {9}},\ \bibinfo {pages} {780} (\bibinfo {year} {2014})}\BibitemShut
  {NoStop}%
\bibitem [{\citenamefont {Gabor}\ \emph {et~al.}(2011)\citenamefont {Gabor},
  \citenamefont {Song}, \citenamefont {Ma}, \citenamefont {Nair}, \citenamefont
  {Taychatanapat}, \citenamefont {Watanabe}, \citenamefont {Taniguchi},
  \citenamefont {Levitov},\ and\ \citenamefont
  {Jarillo-Herrero}}]{Gabor2011HotCarrier}%
  \BibitemOpen
  \bibfield  {author} {\bibinfo {author} {\bibfnamefont {N.~M.}\ \bibnamefont
  {Gabor}}, \bibinfo {author} {\bibfnamefont {J.~C.~W.}\ \bibnamefont {Song}},
  \bibinfo {author} {\bibfnamefont {Q.}~\bibnamefont {Ma}}, \bibinfo {author}
  {\bibfnamefont {N.~L.}\ \bibnamefont {Nair}}, \bibinfo {author}
  {\bibfnamefont {T.}~\bibnamefont {Taychatanapat}}, \bibinfo {author}
  {\bibfnamefont {K.}~\bibnamefont {Watanabe}}, \bibinfo {author}
  {\bibfnamefont {T.}~\bibnamefont {Taniguchi}}, \bibinfo {author}
  {\bibfnamefont {L.~S.}\ \bibnamefont {Levitov}}, \ and\ \bibinfo {author}
  {\bibfnamefont {P.}~\bibnamefont {Jarillo-Herrero}},\ }\href {\doibase
  10.1126/science.1211384} {\bibfield  {journal} {\bibinfo  {journal}
  {Science}\ }\textbf {\bibinfo {volume} {334}},\ \bibinfo {pages} {648}
  (\bibinfo {year} {2011})}\BibitemShut {NoStop}%
\bibitem [{\citenamefont {Jung}\ \emph {et~al.}(2016)\citenamefont {Jung},
  \citenamefont {Rickhaus}, \citenamefont {Zihlmann}, \citenamefont {Makk},\
  and\ \citenamefont {Schönenberger}}]{Jung2016Microwave}%
  \BibitemOpen
  \bibfield  {author} {\bibinfo {author} {\bibfnamefont {M.}~\bibnamefont
  {Jung}}, \bibinfo {author} {\bibfnamefont {P.}~\bibnamefont {Rickhaus}},
  \bibinfo {author} {\bibfnamefont {S.}~\bibnamefont {Zihlmann}}, \bibinfo
  {author} {\bibfnamefont {P.}~\bibnamefont {Makk}}, \ and\ \bibinfo {author}
  {\bibfnamefont {C.}~\bibnamefont {Schönenberger}},\ }\href {\doibase
  10.1021/acs.nanolett.6b03078} {\bibfield  {journal} {\bibinfo  {journal}
  {Nano Lett.}\ }\textbf {\bibinfo {volume} {16}},\ \bibinfo {pages} {6988}
  (\bibinfo {year} {2016})}\BibitemShut {NoStop}%
\bibitem [{\citenamefont {Yan}\ \emph {et~al.}(2012)\citenamefont {Yan},
  \citenamefont {Kim}, \citenamefont {Elle}, \citenamefont {Sushkov},
  \citenamefont {Jenkins}, \citenamefont {Milchberg}, \citenamefont {Fuhrer},\
  and\ \citenamefont {Drew}}]{Fuhrer2012bolometer}%
  \BibitemOpen
  \bibfield  {author} {\bibinfo {author} {\bibfnamefont {J.}~\bibnamefont
  {Yan}}, \bibinfo {author} {\bibfnamefont {M.~H.}\ \bibnamefont {Kim}},
  \bibinfo {author} {\bibfnamefont {J.~A.}\ \bibnamefont {Elle}}, \bibinfo
  {author} {\bibfnamefont {A.~B.}\ \bibnamefont {Sushkov}}, \bibinfo {author}
  {\bibfnamefont {G.~S.}\ \bibnamefont {Jenkins}}, \bibinfo {author}
  {\bibfnamefont {H.~M.}\ \bibnamefont {Milchberg}}, \bibinfo {author}
  {\bibfnamefont {M.~S.}\ \bibnamefont {Fuhrer}}, \ and\ \bibinfo {author}
  {\bibfnamefont {H.}~\bibnamefont {Drew}},\ }\href@noop {} {\bibfield
  {journal} {\bibinfo  {journal} {Nat. Nanotechnol.}\ }\textbf {\bibinfo
  {volume} {7}},\ \bibinfo {pages} {472} (\bibinfo {year} {2012})}\BibitemShut
  {NoStop}%
\bibitem [{\citenamefont {Cai}\ \emph {et~al.}(2014)\citenamefont {Cai},
  \citenamefont {Sushkov}, \citenamefont {Suess}, \citenamefont {Jadidi},
  \citenamefont {Jenkins}, \citenamefont {Nyakiti}, \citenamefont {Myers-Ward},
  \citenamefont {Li}, \citenamefont {Yan}, \citenamefont {Gaskill},
  \citenamefont {Murphy}, \citenamefont {Drew},\ and\ \citenamefont
  {Fuhrer}}]{cai2014sensitive}%
  \BibitemOpen
  \bibfield  {author} {\bibinfo {author} {\bibfnamefont {X.}~\bibnamefont
  {Cai}}, \bibinfo {author} {\bibfnamefont {A.~B.}\ \bibnamefont {Sushkov}},
  \bibinfo {author} {\bibfnamefont {R.~J.}\ \bibnamefont {Suess}}, \bibinfo
  {author} {\bibfnamefont {M.~M.}\ \bibnamefont {Jadidi}}, \bibinfo {author}
  {\bibfnamefont {G.~S.}\ \bibnamefont {Jenkins}}, \bibinfo {author}
  {\bibfnamefont {L.~O.}\ \bibnamefont {Nyakiti}}, \bibinfo {author}
  {\bibfnamefont {R.~L.}\ \bibnamefont {Myers-Ward}}, \bibinfo {author}
  {\bibfnamefont {S.}~\bibnamefont {Li}}, \bibinfo {author} {\bibfnamefont
  {J.}~\bibnamefont {Yan}}, \bibinfo {author} {\bibfnamefont {D.~K.}\
  \bibnamefont {Gaskill}}, \bibinfo {author} {\bibfnamefont {T.~E.}\
  \bibnamefont {Murphy}}, \bibinfo {author} {\bibfnamefont {H.~D.}\
  \bibnamefont {Drew}}, \ and\ \bibinfo {author} {\bibfnamefont {M.~S.}\
  \bibnamefont {Fuhrer}},\ }\href@noop {} {\bibfield  {journal} {\bibinfo
  {journal} {Nat. Nanotechnol.}\ }\textbf {\bibinfo {volume} {9}},\ \bibinfo
  {pages} {814} (\bibinfo {year} {2014})}\BibitemShut {NoStop}%
\bibitem [{\citenamefont {Mittleman}(2013)}]{mittleman2013sensing}%
  \BibitemOpen
  \bibfield  {author} {\bibinfo {author} {\bibfnamefont {D.}~\bibnamefont
  {Mittleman}},\ }\href@noop {} {\emph {\bibinfo {title} {Sensing with
  terahertz radiation}}},\ Vol.~\bibinfo {volume} {85}\ (\bibinfo  {publisher}
  {Springer},\ \bibinfo {year} {2013})\BibitemShut {NoStop}%
\bibitem [{\citenamefont {Vicarelli}\ \emph {et~al.}(2012)\citenamefont
  {Vicarelli}, \citenamefont {Vitiello}, \citenamefont {Coquillat},
  \citenamefont {Lombardo}, \citenamefont {Ferrari}, \citenamefont {Knap},
  \citenamefont {Polini}, \citenamefont {Pellegrini},\ and\ \citenamefont
  {Tredicucci}}]{vicarelli2012graphene}%
  \BibitemOpen
  \bibfield  {author} {\bibinfo {author} {\bibfnamefont {L.}~\bibnamefont
  {Vicarelli}}, \bibinfo {author} {\bibfnamefont {M.}~\bibnamefont {Vitiello}},
  \bibinfo {author} {\bibfnamefont {D.}~\bibnamefont {Coquillat}}, \bibinfo
  {author} {\bibfnamefont {A.}~\bibnamefont {Lombardo}}, \bibinfo {author}
  {\bibfnamefont {A.}~\bibnamefont {Ferrari}}, \bibinfo {author} {\bibfnamefont
  {W.}~\bibnamefont {Knap}}, \bibinfo {author} {\bibfnamefont {M.}~\bibnamefont
  {Polini}}, \bibinfo {author} {\bibfnamefont {V.}~\bibnamefont {Pellegrini}},
  \ and\ \bibinfo {author} {\bibfnamefont {A.}~\bibnamefont {Tredicucci}},\
  }\href@noop {} {\bibfield  {journal} {\bibinfo  {journal} {Nat. Mater.}\
  }\textbf {\bibinfo {volume} {11}},\ \bibinfo {pages} {865} (\bibinfo {year}
  {2012})}\BibitemShut {NoStop}%
\bibitem [{\citenamefont {Spirito}\ \emph {et~al.}(2014)\citenamefont
  {Spirito}, \citenamefont {Coquillat}, \citenamefont {Bonis}, \citenamefont
  {Lombardo}, \citenamefont {Bruna}, \citenamefont {Ferrari}, \citenamefont
  {Pellegrini}, \citenamefont {Tredicucci}, \citenamefont {Knap},\ and\
  \citenamefont {Vitiello}}]{Spirito2014bilayer}%
  \BibitemOpen
  \bibfield  {author} {\bibinfo {author} {\bibfnamefont {D.}~\bibnamefont
  {Spirito}}, \bibinfo {author} {\bibfnamefont {D.}~\bibnamefont {Coquillat}},
  \bibinfo {author} {\bibfnamefont {S.~L.~D.}\ \bibnamefont {Bonis}}, \bibinfo
  {author} {\bibfnamefont {A.}~\bibnamefont {Lombardo}}, \bibinfo {author}
  {\bibfnamefont {M.}~\bibnamefont {Bruna}}, \bibinfo {author} {\bibfnamefont
  {A.~C.}\ \bibnamefont {Ferrari}}, \bibinfo {author} {\bibfnamefont
  {V.}~\bibnamefont {Pellegrini}}, \bibinfo {author} {\bibfnamefont
  {A.}~\bibnamefont {Tredicucci}}, \bibinfo {author} {\bibfnamefont
  {W.}~\bibnamefont {Knap}}, \ and\ \bibinfo {author} {\bibfnamefont {M.~S.}\
  \bibnamefont {Vitiello}},\ }\href {\doibase 10.1063/1.4864082} {\bibfield
  {journal} {\bibinfo  {journal} {Appl. Phys. Lett.}\ }\textbf {\bibinfo
  {volume} {104}},\ \bibinfo {pages} {061111} (\bibinfo {year}
  {2014})}\BibitemShut {NoStop}%
\bibitem [{\citenamefont {Tong}\ \emph {et~al.}(2015)\citenamefont {Tong},
  \citenamefont {Muthee}, \citenamefont {Chen}, \citenamefont {Yngvesson},\
  and\ \citenamefont {Yan}}]{Tong2015AntennaEnhanced}%
  \BibitemOpen
  \bibfield  {author} {\bibinfo {author} {\bibfnamefont {J.}~\bibnamefont
  {Tong}}, \bibinfo {author} {\bibfnamefont {M.}~\bibnamefont {Muthee}},
  \bibinfo {author} {\bibfnamefont {S.-Y.}\ \bibnamefont {Chen}}, \bibinfo
  {author} {\bibfnamefont {S.~K.}\ \bibnamefont {Yngvesson}}, \ and\ \bibinfo
  {author} {\bibfnamefont {J.}~\bibnamefont {Yan}},\ }\href {\doibase
  10.1021/acs.nanolett.5b01635} {\bibfield  {journal} {\bibinfo  {journal}
  {Nano Lett.}\ }\textbf {\bibinfo {volume} {15}},\ \bibinfo {pages} {5295}
  (\bibinfo {year} {2015})}\BibitemShut {NoStop}%
\bibitem [{\citenamefont {Auton}\ \emph {et~al.}(2016)\citenamefont {Auton},
  \citenamefont {Zhang}, \citenamefont {Kumar}, \citenamefont {Wang},
  \citenamefont {Zhang}, \citenamefont {Wang}, \citenamefont {Hill},\ and\
  \citenamefont {Song}}]{auton2016rectifier}%
  \BibitemOpen
  \bibfield  {author} {\bibinfo {author} {\bibfnamefont {G.}~\bibnamefont
  {Auton}}, \bibinfo {author} {\bibfnamefont {J.}~\bibnamefont {Zhang}},
  \bibinfo {author} {\bibfnamefont {R.~K.}\ \bibnamefont {Kumar}}, \bibinfo
  {author} {\bibfnamefont {H.}~\bibnamefont {Wang}}, \bibinfo {author}
  {\bibfnamefont {X.}~\bibnamefont {Zhang}}, \bibinfo {author} {\bibfnamefont
  {Q.}~\bibnamefont {Wang}}, \bibinfo {author} {\bibfnamefont {E.}~\bibnamefont
  {Hill}}, \ and\ \bibinfo {author} {\bibfnamefont {A.}~\bibnamefont {Song}},\
  }\href@noop {} {\bibfield  {journal} {\bibinfo  {journal} {Nat. Commun.}\
  }\textbf {\bibinfo {volume} {7}} (\bibinfo {year} {2016})}\BibitemShut
  {NoStop}%
\bibitem [{\citenamefont {Auton}\ \emph {et~al.}(2017)\citenamefont {Auton},
  \citenamefont {But}, \citenamefont {Zhang}, \citenamefont {Hill},
  \citenamefont {Coquillat}, \citenamefont {Consejo}, \citenamefont {Nouvel},
  \citenamefont {Knap}, \citenamefont {Varani}, \citenamefont {Teppe},
  \citenamefont {Torres},\ and\ \citenamefont
  {Song}}]{Auton2017RectifiersImaging}%
  \BibitemOpen
  \bibfield  {author} {\bibinfo {author} {\bibfnamefont {G.}~\bibnamefont
  {Auton}}, \bibinfo {author} {\bibfnamefont {D.~B.}\ \bibnamefont {But}},
  \bibinfo {author} {\bibfnamefont {J.}~\bibnamefont {Zhang}}, \bibinfo
  {author} {\bibfnamefont {E.}~\bibnamefont {Hill}}, \bibinfo {author}
  {\bibfnamefont {D.}~\bibnamefont {Coquillat}}, \bibinfo {author}
  {\bibfnamefont {C.}~\bibnamefont {Consejo}}, \bibinfo {author} {\bibfnamefont
  {P.}~\bibnamefont {Nouvel}}, \bibinfo {author} {\bibfnamefont
  {W.}~\bibnamefont {Knap}}, \bibinfo {author} {\bibfnamefont {L.}~\bibnamefont
  {Varani}}, \bibinfo {author} {\bibfnamefont {F.}~\bibnamefont {Teppe}},
  \bibinfo {author} {\bibfnamefont {J.}~\bibnamefont {Torres}}, \ and\ \bibinfo
  {author} {\bibfnamefont {A.}~\bibnamefont {Song}},\ }\href {\doibase
  10.1021/acs.nanolett.7b03625} {\bibfield  {journal} {\bibinfo  {journal}
  {Nano Lett.}\ }\textbf {\bibinfo {volume} {17}},\ \bibinfo {pages} {7015}
  (\bibinfo {year} {2017})}\BibitemShut {NoStop}%
\bibitem [{\citenamefont {Generalov}\ \emph {et~al.}(2017)\citenamefont
  {Generalov}, \citenamefont {Andersson}, \citenamefont {Yang}, \citenamefont
  {Vorobiev},\ and\ \citenamefont {Stake}}]{generalov2017400}%
  \BibitemOpen
  \bibfield  {author} {\bibinfo {author} {\bibfnamefont {A.~A.}\ \bibnamefont
  {Generalov}}, \bibinfo {author} {\bibfnamefont {M.~A.}\ \bibnamefont
  {Andersson}}, \bibinfo {author} {\bibfnamefont {X.}~\bibnamefont {Yang}},
  \bibinfo {author} {\bibfnamefont {A.}~\bibnamefont {Vorobiev}}, \ and\
  \bibinfo {author} {\bibfnamefont {J.}~\bibnamefont {Stake}},\ }\href@noop {}
  {\bibfield  {journal} {\bibinfo  {journal} {IEEE Trans. Terahertz Sci.
  Technol.}\ }\textbf {\bibinfo {volume} {7}},\ \bibinfo {pages} {614}
  (\bibinfo {year} {2017})}\BibitemShut {NoStop}%
\bibitem [{\citenamefont {Zak}\ \emph {et~al.}(2014)\citenamefont {Zak},
  \citenamefont {Andersson}, \citenamefont {Bauer}, \citenamefont {Matukas},
  \citenamefont {Lisauskas}, \citenamefont {Roskos},\ and\ \citenamefont
  {Stake}}]{zak2014antenna}%
  \BibitemOpen
  \bibfield  {author} {\bibinfo {author} {\bibfnamefont {A.}~\bibnamefont
  {Zak}}, \bibinfo {author} {\bibfnamefont {M.~A.}\ \bibnamefont {Andersson}},
  \bibinfo {author} {\bibfnamefont {M.}~\bibnamefont {Bauer}}, \bibinfo
  {author} {\bibfnamefont {J.}~\bibnamefont {Matukas}}, \bibinfo {author}
  {\bibfnamefont {A.}~\bibnamefont {Lisauskas}}, \bibinfo {author}
  {\bibfnamefont {H.~G.}\ \bibnamefont {Roskos}}, \ and\ \bibinfo {author}
  {\bibfnamefont {J.}~\bibnamefont {Stake}},\ }\href@noop {} {\bibfield
  {journal} {\bibinfo  {journal} {Nano Lett.}\ }\textbf {\bibinfo {volume}
  {14}},\ \bibinfo {pages} {5834} (\bibinfo {year} {2014})}\BibitemShut
  {NoStop}%
\bibitem [{\citenamefont {Qin}\ \emph {et~al.}(2017)\citenamefont {Qin},
  \citenamefont {Sun}, \citenamefont {Liang}, \citenamefont {Li}, \citenamefont
  {Yang}, \citenamefont {He}, \citenamefont {Yu},\ and\ \citenamefont
  {Feng}}]{qin2017room}%
  \BibitemOpen
  \bibfield  {author} {\bibinfo {author} {\bibfnamefont {H.}~\bibnamefont
  {Qin}}, \bibinfo {author} {\bibfnamefont {J.}~\bibnamefont {Sun}}, \bibinfo
  {author} {\bibfnamefont {S.}~\bibnamefont {Liang}}, \bibinfo {author}
  {\bibfnamefont {X.}~\bibnamefont {Li}}, \bibinfo {author} {\bibfnamefont
  {X.}~\bibnamefont {Yang}}, \bibinfo {author} {\bibfnamefont {Z.}~\bibnamefont
  {He}}, \bibinfo {author} {\bibfnamefont {C.}~\bibnamefont {Yu}}, \ and\
  \bibinfo {author} {\bibfnamefont {Z.}~\bibnamefont {Feng}},\ }\href@noop {}
  {\bibfield  {journal} {\bibinfo  {journal} {Carbon}\ }\textbf {\bibinfo
  {volume} {116}},\ \bibinfo {pages} {760} (\bibinfo {year}
  {2017})}\BibitemShut {NoStop}%
\bibitem [{\citenamefont {Dyakonov}\ and\ \citenamefont
  {Shur}(1996)}]{dyakonov1996detection}%
  \BibitemOpen
  \bibfield  {author} {\bibinfo {author} {\bibfnamefont {M.}~\bibnamefont
  {Dyakonov}}\ and\ \bibinfo {author} {\bibfnamefont {M.}~\bibnamefont
  {Shur}},\ }\href@noop {} {\bibfield  {journal} {\bibinfo  {journal} {IEEE
  Trans. El. Dev.}\ }\textbf {\bibinfo {volume} {43}},\ \bibinfo {pages} {380}
  (\bibinfo {year} {1996})}\BibitemShut {NoStop}%
\bibitem [{\citenamefont {Tomadin}\ and\ \citenamefont
  {Polini}(2013)}]{Tomadin2013Plasma}%
  \BibitemOpen
  \bibfield  {author} {\bibinfo {author} {\bibfnamefont {A.}~\bibnamefont
  {Tomadin}}\ and\ \bibinfo {author} {\bibfnamefont {M.}~\bibnamefont
  {Polini}},\ }\href {\doibase 10.1103/PhysRevB.88.205426} {\bibfield
  {journal} {\bibinfo  {journal} {Phys. Rev. B}\ }\textbf {\bibinfo {volume}
  {88}},\ \bibinfo {pages} {205426} (\bibinfo {year} {2013})}\BibitemShut
  {NoStop}%
\bibitem [{\citenamefont {Kretinin}\ \emph {et~al.}(2014)\citenamefont
  {Kretinin}, \citenamefont {Cao}, \citenamefont {Tu}, \citenamefont {Yu},
  \citenamefont {Jalil}, \citenamefont {Novoselov}, \citenamefont {Haigh},
  \citenamefont {Gholinia}, \citenamefont {Mishchenko}, \citenamefont {Lozada},
  \citenamefont {Georgiou}, \citenamefont {Woods}, \citenamefont {Withers},
  \citenamefont {Blake}, \citenamefont {Eda}, \citenamefont {Wirsig},
  \citenamefont {Hucho}, \citenamefont {Watanabe}, \citenamefont {Taniguchi},
  \citenamefont {Geim},\ and\ \citenamefont
  {Gorbachev}}]{kretinin2014electronic}%
  \BibitemOpen
  \bibfield  {author} {\bibinfo {author} {\bibfnamefont {A.}~\bibnamefont
  {Kretinin}}, \bibinfo {author} {\bibfnamefont {Y.}~\bibnamefont {Cao}},
  \bibinfo {author} {\bibfnamefont {J.}~\bibnamefont {Tu}}, \bibinfo {author}
  {\bibfnamefont {G.}~\bibnamefont {Yu}}, \bibinfo {author} {\bibfnamefont
  {R.}~\bibnamefont {Jalil}}, \bibinfo {author} {\bibfnamefont
  {K.}~\bibnamefont {Novoselov}}, \bibinfo {author} {\bibfnamefont
  {S.}~\bibnamefont {Haigh}}, \bibinfo {author} {\bibfnamefont
  {A.}~\bibnamefont {Gholinia}}, \bibinfo {author} {\bibfnamefont
  {A.}~\bibnamefont {Mishchenko}}, \bibinfo {author} {\bibfnamefont
  {M.}~\bibnamefont {Lozada}}, \bibinfo {author} {\bibfnamefont
  {T.}~\bibnamefont {Georgiou}}, \bibinfo {author} {\bibfnamefont
  {C.}~\bibnamefont {Woods}}, \bibinfo {author} {\bibfnamefont
  {F.}~\bibnamefont {Withers}}, \bibinfo {author} {\bibfnamefont
  {P.}~\bibnamefont {Blake}}, \bibinfo {author} {\bibfnamefont
  {G.}~\bibnamefont {Eda}}, \bibinfo {author} {\bibfnamefont {A.}~\bibnamefont
  {Wirsig}}, \bibinfo {author} {\bibfnamefont {C.}~\bibnamefont {Hucho}},
  \bibinfo {author} {\bibfnamefont {K.}~\bibnamefont {Watanabe}}, \bibinfo
  {author} {\bibfnamefont {T.}~\bibnamefont {Taniguchi}}, \bibinfo {author}
  {\bibfnamefont {A.}~\bibnamefont {Geim}}, \ and\ \bibinfo {author}
  {\bibfnamefont {R.}~\bibnamefont {Gorbachev}},\ }\href@noop {} {\bibfield
  {journal} {\bibinfo  {journal} {Nano Lett.}\ }\textbf {\bibinfo {volume}
  {14}},\ \bibinfo {pages} {3270} (\bibinfo {year} {2014})}\BibitemShut
  {NoStop}%
\bibitem [{\citenamefont {Wang}\ \emph {et~al.}(2013)\citenamefont {Wang},
  \citenamefont {Meric}, \citenamefont {Huang}, \citenamefont {Gao},
  \citenamefont {Gao}, \citenamefont {Tran}, \citenamefont {Taniguchi},
  \citenamefont {Watanabe}, \citenamefont {Campos}, \citenamefont {Muller},
  \citenamefont {Guo}, \citenamefont {Kim}, \citenamefont {Hone}, \citenamefont
  {Shepard},\ and\ \citenamefont {Dean}}]{Wang2013onedimensional}%
  \BibitemOpen
  \bibfield  {author} {\bibinfo {author} {\bibfnamefont {L.}~\bibnamefont
  {Wang}}, \bibinfo {author} {\bibfnamefont {I.}~\bibnamefont {Meric}},
  \bibinfo {author} {\bibfnamefont {P.~Y.}\ \bibnamefont {Huang}}, \bibinfo
  {author} {\bibfnamefont {Q.}~\bibnamefont {Gao}}, \bibinfo {author}
  {\bibfnamefont {Y.}~\bibnamefont {Gao}}, \bibinfo {author} {\bibfnamefont
  {H.}~\bibnamefont {Tran}}, \bibinfo {author} {\bibfnamefont {T.}~\bibnamefont
  {Taniguchi}}, \bibinfo {author} {\bibfnamefont {K.}~\bibnamefont {Watanabe}},
  \bibinfo {author} {\bibfnamefont {L.~M.}\ \bibnamefont {Campos}}, \bibinfo
  {author} {\bibfnamefont {D.~A.}\ \bibnamefont {Muller}}, \bibinfo {author}
  {\bibfnamefont {J.}~\bibnamefont {Guo}}, \bibinfo {author} {\bibfnamefont
  {P.}~\bibnamefont {Kim}}, \bibinfo {author} {\bibfnamefont {J.}~\bibnamefont
  {Hone}}, \bibinfo {author} {\bibfnamefont {K.~L.}\ \bibnamefont {Shepard}}, \
  and\ \bibinfo {author} {\bibfnamefont {C.~R.}\ \bibnamefont {Dean}},\ }\href
  {\doibase 10.1126/science.1244358} {\bibfield  {journal} {\bibinfo  {journal}
  {Science}\ }\textbf {\bibinfo {volume} {342}},\ \bibinfo {pages} {614}
  (\bibinfo {year} {2013})}\BibitemShut {NoStop}%
\bibitem [{\citenamefont {Fedorov}\ \emph {et~al.}(2013)\citenamefont
  {Fedorov}, \citenamefont {Kardakova}, \citenamefont {Gayduchenko},
  \citenamefont {Charayev}, \citenamefont {Voronov}, \citenamefont {Finkel},
  \citenamefont {Klapwijk}, \citenamefont {Morozov}, \citenamefont
  {Presniakov}, \citenamefont {Bobrinetskiy}, \citenamefont {Ibragimov},\ and\
  \citenamefont {Goltsman}}]{fedorov2013photothermoelectric}%
  \BibitemOpen
  \bibfield  {author} {\bibinfo {author} {\bibfnamefont {G.}~\bibnamefont
  {Fedorov}}, \bibinfo {author} {\bibfnamefont {A.}~\bibnamefont {Kardakova}},
  \bibinfo {author} {\bibfnamefont {I.}~\bibnamefont {Gayduchenko}}, \bibinfo
  {author} {\bibfnamefont {I.}~\bibnamefont {Charayev}}, \bibinfo {author}
  {\bibfnamefont {B.~M.}\ \bibnamefont {Voronov}}, \bibinfo {author}
  {\bibfnamefont {M.}~\bibnamefont {Finkel}}, \bibinfo {author} {\bibfnamefont
  {T.~M.}\ \bibnamefont {Klapwijk}}, \bibinfo {author} {\bibfnamefont
  {S.}~\bibnamefont {Morozov}}, \bibinfo {author} {\bibfnamefont
  {M.}~\bibnamefont {Presniakov}}, \bibinfo {author} {\bibfnamefont
  {I.}~\bibnamefont {Bobrinetskiy}}, \bibinfo {author} {\bibfnamefont
  {R.}~\bibnamefont {Ibragimov}}, \ and\ \bibinfo {author} {\bibfnamefont
  {G.}~\bibnamefont {Goltsman}},\ }\href {\doibase 10.1063/1.4828555}
  {\bibfield  {journal} {\bibinfo  {journal} {Applied Physics Letters}\
  }\textbf {\bibinfo {volume} {103}},\ \bibinfo {pages} {181121} (\bibinfo
  {year} {2013})}\BibitemShut {NoStop}%
\bibitem [{\citenamefont {Vasilyev}\ \emph {et~al.}(2017)\citenamefont
  {Vasilyev}, \citenamefont {Vasileva}, \citenamefont {Novikov}, \citenamefont
  {Tarasenko}, \citenamefont {Danilov},\ and\ \citenamefont
  {Ganichev}}]{vasilyev2017high}%
  \BibitemOpen
  \bibfield  {author} {\bibinfo {author} {\bibfnamefont {Y.~B.}\ \bibnamefont
  {Vasilyev}}, \bibinfo {author} {\bibfnamefont {G.~Y.}\ \bibnamefont
  {Vasileva}}, \bibinfo {author} {\bibfnamefont {S.}~\bibnamefont {Novikov}},
  \bibinfo {author} {\bibfnamefont {S.}~\bibnamefont {Tarasenko}}, \bibinfo
  {author} {\bibfnamefont {S.}~\bibnamefont {Danilov}}, \ and\ \bibinfo
  {author} {\bibfnamefont {S.}~\bibnamefont {Ganichev}},\ }\href@noop {}
  {\bibfield  {journal} {\bibinfo  {journal} {arXiv preprint arXiv:1711.03803}\
  } (\bibinfo {year} {2017})}\BibitemShut {NoStop}%
\bibitem [{\citenamefont {Nouchi}, \citenamefont {Saito},\ and\ \citenamefont
  {Tanigaki}(2011)}]{nouchi2011determination}%
  \BibitemOpen
  \bibfield  {author} {\bibinfo {author} {\bibfnamefont {R.}~\bibnamefont
  {Nouchi}}, \bibinfo {author} {\bibfnamefont {T.}~\bibnamefont {Saito}}, \
  and\ \bibinfo {author} {\bibfnamefont {K.}~\bibnamefont {Tanigaki}},\
  }\href@noop {} {\bibfield  {journal} {\bibinfo  {journal} {Appl. Phys.
  Express}\ }\textbf {\bibinfo {volume} {4}},\ \bibinfo {pages} {035101}
  (\bibinfo {year} {2011})}\BibitemShut {NoStop}%
\bibitem [{\citenamefont {McCreary}\ \emph {et~al.}(2010)\citenamefont
  {McCreary}, \citenamefont {Pi}, \citenamefont {Swartz}, \citenamefont {Han},
  \citenamefont {Bao}, \citenamefont {Lau}, \citenamefont {Guinea},
  \citenamefont {Katsnelson},\ and\ \citenamefont
  {Kawakami}}]{mccreary2010effect}%
  \BibitemOpen
  \bibfield  {author} {\bibinfo {author} {\bibfnamefont {K.}~\bibnamefont
  {McCreary}}, \bibinfo {author} {\bibfnamefont {K.}~\bibnamefont {Pi}},
  \bibinfo {author} {\bibfnamefont {A.}~\bibnamefont {Swartz}}, \bibinfo
  {author} {\bibfnamefont {W.}~\bibnamefont {Han}}, \bibinfo {author}
  {\bibfnamefont {W.}~\bibnamefont {Bao}}, \bibinfo {author} {\bibfnamefont
  {C.}~\bibnamefont {Lau}}, \bibinfo {author} {\bibfnamefont {F.}~\bibnamefont
  {Guinea}}, \bibinfo {author} {\bibfnamefont {M.}~\bibnamefont {Katsnelson}},
  \ and\ \bibinfo {author} {\bibfnamefont {R.}~\bibnamefont {Kawakami}},\
  }\href@noop {} {\bibfield  {journal} {\bibinfo  {journal} {Phys. Rev. B}\
  }\textbf {\bibinfo {volume} {81}},\ \bibinfo {pages} {115453} (\bibinfo
  {year} {2010})}\BibitemShut {NoStop}%
\bibitem [{\citenamefont {Lee}\ \emph {et~al.}(2008)\citenamefont {Lee},
  \citenamefont {Balasubramanian}, \citenamefont {Weitz}, \citenamefont
  {Burghard},\ and\ \citenamefont {Kern}}]{lee2008contact}%
  \BibitemOpen
  \bibfield  {author} {\bibinfo {author} {\bibfnamefont {E.~J.}\ \bibnamefont
  {Lee}}, \bibinfo {author} {\bibfnamefont {K.}~\bibnamefont
  {Balasubramanian}}, \bibinfo {author} {\bibfnamefont {R.~T.}\ \bibnamefont
  {Weitz}}, \bibinfo {author} {\bibfnamefont {M.}~\bibnamefont {Burghard}}, \
  and\ \bibinfo {author} {\bibfnamefont {K.}~\bibnamefont {Kern}},\ }\href@noop
  {} {\bibfield  {journal} {\bibinfo  {journal} {Nat. Nanotechnol.}\ }\textbf
  {\bibinfo {volume} {3}},\ \bibinfo {pages} {486} (\bibinfo {year}
  {2008})}\BibitemShut {NoStop}%
\bibitem [{\citenamefont {Bandurin}\ \emph {et~al.}(2016)\citenamefont
  {Bandurin}, \citenamefont {Torre}, \citenamefont {Kumar}, \citenamefont
  {Ben~Shalom}, \citenamefont {Tomadin}, \citenamefont {Principi},
  \citenamefont {Auton}, \citenamefont {Khestanova}, \citenamefont {Novoselov},
  \citenamefont {Grigorieva}, \citenamefont {Ponomarenko}, \citenamefont
  {Geim},\ and\ \citenamefont {Polini}}]{Bandurin-HDgraphene}%
  \BibitemOpen
  \bibfield  {author} {\bibinfo {author} {\bibfnamefont {D.~A.}\ \bibnamefont
  {Bandurin}}, \bibinfo {author} {\bibfnamefont {I.}~\bibnamefont {Torre}},
  \bibinfo {author} {\bibfnamefont {R.~K.}\ \bibnamefont {Kumar}}, \bibinfo
  {author} {\bibfnamefont {M.}~\bibnamefont {Ben~Shalom}}, \bibinfo {author}
  {\bibfnamefont {A.}~\bibnamefont {Tomadin}}, \bibinfo {author} {\bibfnamefont
  {A.}~\bibnamefont {Principi}}, \bibinfo {author} {\bibfnamefont {G.~H.}\
  \bibnamefont {Auton}}, \bibinfo {author} {\bibfnamefont {E.}~\bibnamefont
  {Khestanova}}, \bibinfo {author} {\bibfnamefont {K.~S.}\ \bibnamefont
  {Novoselov}}, \bibinfo {author} {\bibfnamefont {I.~V.}\ \bibnamefont
  {Grigorieva}}, \bibinfo {author} {\bibfnamefont {L.~A.}\ \bibnamefont
  {Ponomarenko}}, \bibinfo {author} {\bibfnamefont {A.~K.}\ \bibnamefont
  {Geim}}, \ and\ \bibinfo {author} {\bibfnamefont {M.}~\bibnamefont
  {Polini}},\ }\href {\doibase 10.1126/science.aad0201} {\bibfield  {journal}
  {\bibinfo  {journal} {Science}\ }\textbf {\bibinfo {volume} {351}},\ \bibinfo
  {pages} {1055} (\bibinfo {year} {2016})}\BibitemShut {NoStop}%
\bibitem [{\citenamefont {Allen}\ \emph {et~al.}(2015)\citenamefont {Allen},
  \citenamefont {Shtanko}, \citenamefont {Fulga}, \citenamefont {Wang},
  \citenamefont {Nurgaliev}, \citenamefont {Watanabe}, \citenamefont
  {Taniguchi}, \citenamefont {Akhmerov}, \citenamefont {Jarillo-Herrero},
  \citenamefont {Levitov},\ and\ \citenamefont
  {Yacoby}}]{allen2015visualization}%
  \BibitemOpen
  \bibfield  {author} {\bibinfo {author} {\bibfnamefont {M.}~\bibnamefont
  {Allen}}, \bibinfo {author} {\bibfnamefont {O.}~\bibnamefont {Shtanko}},
  \bibinfo {author} {\bibfnamefont {I.~C.}\ \bibnamefont {Fulga}}, \bibinfo
  {author} {\bibfnamefont {J.-J.}\ \bibnamefont {Wang}}, \bibinfo {author}
  {\bibfnamefont {D.}~\bibnamefont {Nurgaliev}}, \bibinfo {author}
  {\bibfnamefont {K.}~\bibnamefont {Watanabe}}, \bibinfo {author}
  {\bibfnamefont {T.}~\bibnamefont {Taniguchi}}, \bibinfo {author}
  {\bibfnamefont {A.}~\bibnamefont {Akhmerov}}, \bibinfo {author}
  {\bibfnamefont {P.}~\bibnamefont {Jarillo-Herrero}}, \bibinfo {author}
  {\bibfnamefont {L.}~\bibnamefont {Levitov}}, \ and\ \bibinfo {author}
  {\bibfnamefont {A.}~\bibnamefont {Yacoby}},\ }\href@noop {} {\bibfield
  {journal} {\bibinfo  {journal} {arXiv preprint arXiv:1506.06734}\ } (\bibinfo
  {year} {2015})}\BibitemShut {NoStop}%
\bibitem [{\citenamefont {Shalom}\ \emph {et~al.}(2016)\citenamefont {Shalom},
  \citenamefont {Zhu}, \citenamefont {Fal’ko}, \citenamefont {Mishchenko},
  \citenamefont {Kretinin}, \citenamefont {Novoselov}, \citenamefont {Woods},
  \citenamefont {Watanabe}, \citenamefont {Taniguchi}, \citenamefont {Geim},\
  and\ \citenamefont {Prance}}]{shalom2016quantum}%
  \BibitemOpen
  \bibfield  {author} {\bibinfo {author} {\bibfnamefont {M.~B.}\ \bibnamefont
  {Shalom}}, \bibinfo {author} {\bibfnamefont {M.}~\bibnamefont {Zhu}},
  \bibinfo {author} {\bibfnamefont {V.}~\bibnamefont {Fal’ko}}, \bibinfo
  {author} {\bibfnamefont {A.}~\bibnamefont {Mishchenko}}, \bibinfo {author}
  {\bibfnamefont {A.}~\bibnamefont {Kretinin}}, \bibinfo {author}
  {\bibfnamefont {K.}~\bibnamefont {Novoselov}}, \bibinfo {author}
  {\bibfnamefont {C.}~\bibnamefont {Woods}}, \bibinfo {author} {\bibfnamefont
  {K.}~\bibnamefont {Watanabe}}, \bibinfo {author} {\bibfnamefont
  {T.}~\bibnamefont {Taniguchi}}, \bibinfo {author} {\bibfnamefont
  {A.}~\bibnamefont {Geim}}, \ and\ \bibinfo {author} {\bibfnamefont
  {J.}~\bibnamefont {Prance}},\ }\href@noop {} {\bibfield  {journal} {\bibinfo
  {journal} {Nat. Phys.}\ }\textbf {\bibinfo {volume} {12}},\ \bibinfo {pages}
  {318} (\bibinfo {year} {2016})}\BibitemShut {NoStop}%
\bibitem [{\citenamefont {Xu}\ \emph {et~al.}(2009)\citenamefont {Xu},
  \citenamefont {Gabor}, \citenamefont {Alden}, \citenamefont {van~der Zande},\
  and\ \citenamefont {McEuen}}]{xu2009photo}%
  \BibitemOpen
  \bibfield  {author} {\bibinfo {author} {\bibfnamefont {X.}~\bibnamefont
  {Xu}}, \bibinfo {author} {\bibfnamefont {N.~M.}\ \bibnamefont {Gabor}},
  \bibinfo {author} {\bibfnamefont {J.~S.}\ \bibnamefont {Alden}}, \bibinfo
  {author} {\bibfnamefont {A.~M.}\ \bibnamefont {van~der Zande}}, \ and\
  \bibinfo {author} {\bibfnamefont {P.~L.}\ \bibnamefont {McEuen}},\
  }\href@noop {} {\bibfield  {journal} {\bibinfo  {journal} {Nano Lett.}\
  }\textbf {\bibinfo {volume} {10}},\ \bibinfo {pages} {562} (\bibinfo {year}
  {2009})}\BibitemShut {NoStop}%
\bibitem [{\citenamefont {Crossno}\ \emph {et~al.}(2015)\citenamefont
  {Crossno}, \citenamefont {Liu}, \citenamefont {Ohki}, \citenamefont {Kim},\
  and\ \citenamefont {Fong}}]{crossno2015thermometry}%
  \BibitemOpen
  \bibfield  {author} {\bibinfo {author} {\bibfnamefont {J.}~\bibnamefont
  {Crossno}}, \bibinfo {author} {\bibfnamefont {X.}~\bibnamefont {Liu}},
  \bibinfo {author} {\bibfnamefont {T.~A.}\ \bibnamefont {Ohki}}, \bibinfo
  {author} {\bibfnamefont {P.}~\bibnamefont {Kim}}, \ and\ \bibinfo {author}
  {\bibfnamefont {K.~C.}\ \bibnamefont {Fong}},\ }\href@noop {} {\bibfield
  {journal} {\bibinfo  {journal} {Appl. Phys. Lett.}\ }\textbf {\bibinfo
  {volume} {106}},\ \bibinfo {pages} {023121} (\bibinfo {year}
  {2015})}\BibitemShut {NoStop}%
\bibitem [{\citenamefont {Fong}\ \emph {et~al.}(2013)\citenamefont {Fong},
  \citenamefont {Wollman}, \citenamefont {Ravi}, \citenamefont {Chen},
  \citenamefont {Clerk}, \citenamefont {Shaw}, \citenamefont {Leduc},\ and\
  \citenamefont {Schwab}}]{Fong2013}%
  \BibitemOpen
  \bibfield  {author} {\bibinfo {author} {\bibfnamefont {K.~C.}\ \bibnamefont
  {Fong}}, \bibinfo {author} {\bibfnamefont {E.~E.}\ \bibnamefont {Wollman}},
  \bibinfo {author} {\bibfnamefont {H.}~\bibnamefont {Ravi}}, \bibinfo {author}
  {\bibfnamefont {W.}~\bibnamefont {Chen}}, \bibinfo {author} {\bibfnamefont
  {A.~A.}\ \bibnamefont {Clerk}}, \bibinfo {author} {\bibfnamefont {M.~D.}\
  \bibnamefont {Shaw}}, \bibinfo {author} {\bibfnamefont {H.~G.}\ \bibnamefont
  {Leduc}}, \ and\ \bibinfo {author} {\bibfnamefont {K.~C.}\ \bibnamefont
  {Schwab}},\ }\href {\doibase 10.1103/PhysRevX.3.041008} {\bibfield  {journal}
  {\bibinfo  {journal} {Phys. Rev. X}\ }\textbf {\bibinfo {volume} {3}},\
  \bibinfo {pages} {041008} (\bibinfo {year} {2013})}\BibitemShut {NoStop}%
\bibitem [{\citenamefont {Tielrooij}\ \emph {et~al.}(2017)\citenamefont
  {Tielrooij}, \citenamefont {Hesp}, \citenamefont {Principi}, \citenamefont
  {Lundeberg}, \citenamefont {Pogna}, \citenamefont {Banszerus}, \citenamefont
  {Mics}, \citenamefont {Massicotte}, \citenamefont {Schmidt}, \citenamefont
  {Davydovskaya} \emph {et~al.}}]{tielrooij2017out}%
  \BibitemOpen
  \bibfield  {author} {\bibinfo {author} {\bibfnamefont {K.}~\bibnamefont
  {Tielrooij}}, \bibinfo {author} {\bibfnamefont {N.}~\bibnamefont {Hesp}},
  \bibinfo {author} {\bibfnamefont {A.}~\bibnamefont {Principi}}, \bibinfo
  {author} {\bibfnamefont {M.}~\bibnamefont {Lundeberg}}, \bibinfo {author}
  {\bibfnamefont {E.}~\bibnamefont {Pogna}}, \bibinfo {author} {\bibfnamefont
  {L.}~\bibnamefont {Banszerus}}, \bibinfo {author} {\bibfnamefont
  {Z.}~\bibnamefont {Mics}}, \bibinfo {author} {\bibfnamefont {M.}~\bibnamefont
  {Massicotte}}, \bibinfo {author} {\bibfnamefont {P.}~\bibnamefont {Schmidt}},
  \bibinfo {author} {\bibfnamefont {D.}~\bibnamefont {Davydovskaya}},  \emph
  {et~al.},\ }\href@noop {} {\bibfield  {journal} {\bibinfo  {journal} {arXiv
  preprint arXiv:1702.03766}\ } (\bibinfo {year} {2017})}\BibitemShut {NoStop}%
\bibitem [{\citenamefont {Shurakov}\ \emph {et~al.}(2012)\citenamefont
  {Shurakov}, \citenamefont {Seliverstov}, \citenamefont {Kaurova},
  \citenamefont {Finkel}, \citenamefont {Voronov},\ and\ \citenamefont
  {Goltsman}}]{Shurakov2012Antenna}%
  \BibitemOpen
  \bibfield  {author} {\bibinfo {author} {\bibfnamefont {A.}~\bibnamefont
  {Shurakov}}, \bibinfo {author} {\bibfnamefont {S.}~\bibnamefont
  {Seliverstov}}, \bibinfo {author} {\bibfnamefont {N.}~\bibnamefont
  {Kaurova}}, \bibinfo {author} {\bibfnamefont {M.}~\bibnamefont {Finkel}},
  \bibinfo {author} {\bibfnamefont {B.}~\bibnamefont {Voronov}}, \ and\
  \bibinfo {author} {\bibfnamefont {G.}~\bibnamefont {Goltsman}},\ }\href
  {\doibase 10.1109/TTHZ.2012.2194852} {\bibfield  {journal} {\bibinfo
  {journal} {IEEE Transactions on Terahertz Science and Technology}\ }\textbf
  {\bibinfo {volume} {2}},\ \bibinfo {pages} {400} (\bibinfo {year}
  {2012})}\BibitemShut {NoStop}%
\bibitem [{\citenamefont {Milligan}(2005)}]{milligan2005modern}%
  \BibitemOpen
  \bibfield  {author} {\bibinfo {author} {\bibfnamefont {T.~A.}\ \bibnamefont
  {Milligan}},\ }\href@noop {} {\emph {\bibinfo {title} {Modern antenna
  design}}}\ (\bibinfo  {publisher} {John Wiley \& Sons},\ \bibinfo {year}
  {2005})\BibitemShut {NoStop}%
\end{thebibliography}%

\begin{widetext}
\setcounter{equation}{0}
\setcounter{figure}{0}
\renewcommand{\theequation} {S\arabic{equation}}
\renewcommand{\thefigure} {S\arabic{figure}}
\section*{Supporting information}
\section{Sample fabrication}

Our graphene-based FET was fabricated by dry-transfer technique as described elsewhere~\cite{kretinin2014electronic}. This involved mechanical exfoliation to obtain single layer graphene (SLG) and hexagonal boron nitride (hBN) crystals ($\approx$50 nm thick). The flakes were stacked on top of each other using polymer membrane and deposited on top of an oxidized (500 nm of SiO$_2$) boron$-–$doped silicon wafer. The low conductivity of the wafer ensured its transparency to the THz and sub-THz radiation but it was high enough to use it as a back gate. Afterwards, the obtained hBN/SLG/hBN heterostructure was patterned using electron beam lithography to define quasi-one-dimensional contacts to graphene (3nm of chromium, 50 nm of gold). The next round of e-beam lithography was used to define a narrow top gate of 500 nm in width. After that, a PMMA mask (L=6 $\mu$m x $W=2.6$ $\mu$m) was lithographically patterned, and reactive ion etching was used to translate the shape of the mask to the heterostructure. Finally, optical lithography was used to extend the source and the top gate electrodes to the micrometer scale and pattern them in the shape of a broadband logarithmic spiral antenna (5 nm of Ti, 200 nm of Au), Fig. S1. 

The choice of the antenna is defined both by simplicity of its fabrication and broadband characteristics. For our experiments  we have chosen the log-spiral  antenna defined in polar coordinates as   $R={{R}_{0}}{{e}^{\varphi/b }}$ with following parameters: the inner radius of the spiral  ${{R}_{0}}$ equals 5.5 $\mu$m, outer radius  $R_{\max} = 68$ $\mu$m, the parameter determining the rate of spiral $b=3.2$. Further details of the antenna parameters and its broadband characteristics are presented in Ref. [\onlinecite{Shurakov2012Antenna}]. \addDS{We note here that the contact to the drain is intentionally made thin and closely adjacent to the gate sleeve. This creates the asymmetry required for THz detection, i.e. the electromotive force under illumination develops dominantly between source and gate terminals.}

\begin{figure}[h]
	\centering\includegraphics[width=0.3\linewidth]{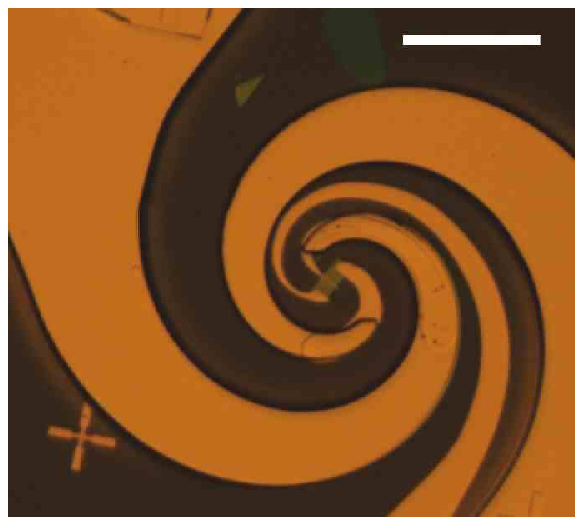}
	\caption{Optical photograph of the detector antenna. Scale bar is 30 $\mu$m.}
\end{figure}

\section{Broadband photoresponse and noise equivalent power}

In Fig. S2a we plot examples of room temperature responsivity $R_a (V_{bg})$ acquired at different frequencies of incoming radiation. The  device exhibits a broadband photoresponse at all gate voltages away from the CNP.

The characterization of any photodetector requires the measurements of its noise equivalent power NEP. The latter is defined as a ratio between the noise spectral density $S_V$ and the photoresponsivity $NEP=S_V/(|R_a |)$. The dominant source of noise in FETs is the Johnson$-$Nyquist contribution with $S_V=(4k_B TR)^{1/2}$ where $k_B$ is the Boltzmann constant. In Fig. S2b we plot $NEP(V_{bg})$ \addDB{calculated using Johnson$-$Nyquist law for the resistance of our FET. } The minimum $NEP=0.6$ nW/Hz$^{1/2}$ detected in our device is comparable to that reported previously for non-encapsulated devices on $SiO_2$~\cite{vicarelli2012graphene,Spirito2014bilayer,qin2017room}. Further reduction of the NEP can be achieved by lowering graphene-metal contact resistance. 

\begin{figure}[h]
\centering\includegraphics[width=0.8\linewidth]{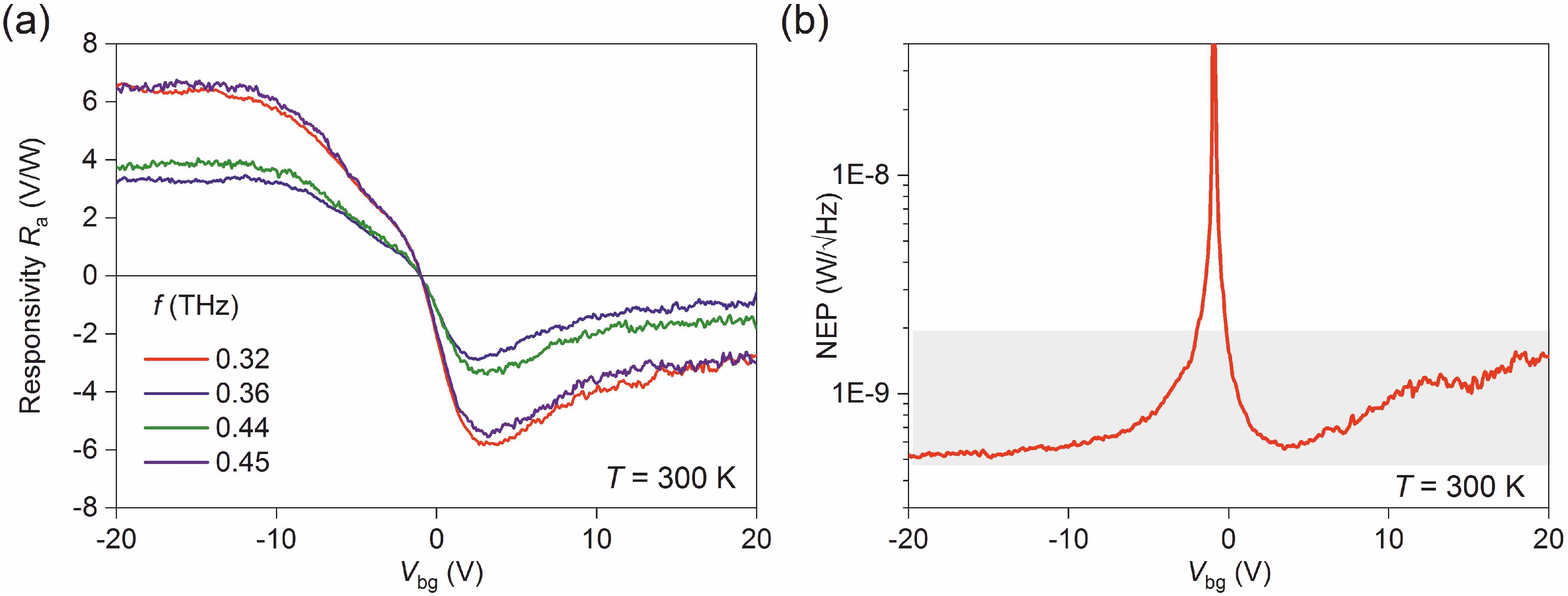}
\label{fig-broadband}
\caption{ (a) $R_a$ as a function of $V_{bg}$ for different frequencies of incoming radiation. (b) RT NEP as a function of $V_{bg}$ obtained at $f=0.13$ Thz.}
\end{figure}

\section{Calculation of photo-thermoelectric response}
In this section, we discuss the thermo-photovoltaic voltage and its dependence on Seebeck coefficient and thermal conductivity in a graphene-based FET designed in the configuration shown in Fig. S3. For this purpose, we solve the heat conduction equation for electrons
\begin{gather}
\label{eq-heat-conduction}
q\left( x \right)=-\frac{\partial }{\partial x}\chi \frac{\partial T}{\partial x} - C\frac{T-{{T}_{0}}}{{{\tau }_{\varepsilon }}}, \\ 
T\left( 0 \right)=T\left( L \right)={{T}_{0}}, 
\end{gather}
where $\chi$ is the thermal conductivity, $q\left( x \right)$ is the dissipated heat per unit area. In the case of Joule heating the dissipated heat and the electric field are related as $q\left( x \right)=\operatorname{Re}\sigma {{\left| E\left( x \right) \right|}^{2}}/2,$ where $\sigma$ is the conductivity and $E\left( x \right)$ is the longitudinal electric field in the channel. The last term on the right-hand side describes heat sink to phonons, for simplicity of interpretation, we have divided the prefactor in this coefficient into heat capacity $C$ and energy relaxation time ${{\tau }_{\varepsilon }}$. We have assumed that the phonon temperature is constant and equal to the electron temperature at the contacts $x=0, L$.

\begin{figure}[h]
\centering\includegraphics[width=0.6\linewidth]{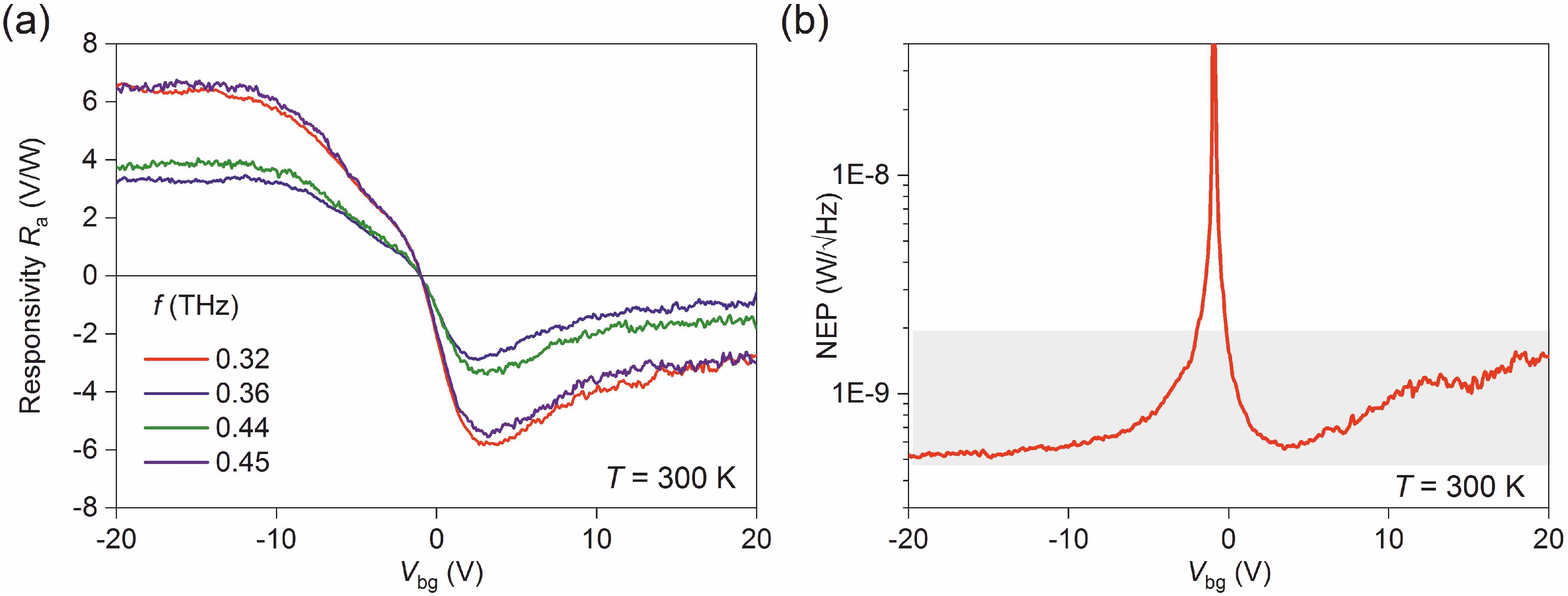}
\label{figure-fet-scheme}
\caption{Schematic of a dual-gated field effect transistor. Orange – graphene channel, blue – doped graphene sheet in the vicinity of contacts, green – hexagonal boron nitride, grey – top gate}
\end{figure}

For a simple analytical solution, we assume that only the left part of our device is heated by the ac current. This is because of the open-circuit condition at the drain (see next section); the electrical current injected from the source current drains as a displacement current into the top gate capacitor. Therefore, $q\left( x \right)=\theta \left( L/2-x \right)\operatorname{Re}\sigma {{\left| {{E}_{0}} \right|}^{2}}/2$, where $E_0$ is the coordinate-independent electric field in the channel, and $\theta(x)$ is the Heaviside theta function.  After a simple scaling $\xi =x/L$, the eq.~(\ref{eq-heat-conduction}) becomes
\begin{gather}
\frac{{{\partial }^{2}}\delta T}{\partial {{\xi }^{2}}}+{{\left( \frac{L}{{{L}_{T}}} \right)}^{2}}\delta T=-\frac{{{L}^{2}}\operatorname{Re}\sigma E_{0}^{2}}{2\chi }\theta \left( 1/2-\xi  \right), \\ 
\,\delta T\left( 0 \right)=\delta T\left( L \right)=0, 
\end{gather}
where we have introduced the thermal relaxation length ${L_T}=\sqrt{\chi {\tau_{\varepsilon }}/C}$. Its solution is a linear combination of growing and decaying exponentials and is written straightforwardly.

The thermovoltage of the device shown in Fig. S2 reads
\begin{equation}
\label{Eq-UPTE}
\Delta {U_{\rm PTE}}=-\int\limits_{{{T}_{0}}}^{{{T}_{S}}}{{{S}_{cont}}dT}-\int\limits_{{{T}_{S}}}^{{{T}_{D}}}{{{S}_{ch}}dT}-\int\limits_{{{T}_{D}}}^{{{T}_{0}}}{{{S}_{cont}}dT}=\left( {{S}_{ch}}-{{S}_{cont}} \right)\left( {{T}_{S}}-{{T}_{D}} \right),
\end{equation}
where $T_0$ is the temperature of the metal contacts, $T_S$ ($T_D$) is the temperature at the boundaries between the left (right) doped areas and the graphene channel. The temperature difference between the near-contact regions can be expressed as
\begin{equation}
{{T}_{S}}-{{T}_{D}}={{\left. \delta T \right|}_{x=\delta L}}-{{\left. \delta T \right|}_{x=L-\delta L}}.
\end{equation}
Using the smallness of the contact length $\delta L \ll L$, we can write
	\[{{\left. \delta T \right|}_{x=\delta L}}-{{\left. \delta T \right|}_{x=L-\delta L}}\approx \left( {{\left. \frac{\partial \delta T}{\partial x} \right|}_{x=0}}+{{\left. \frac{\partial \delta T}{\partial x} \right|}_{x=L}} \right)\delta L\]
Substituting the solution of eq.~\ref{eq-heat-conduction} into \ref{Eq-UPTE}, we obtain
\begin{equation}
   {{T}_{S}}-{{T}_{D}}=\frac{\operatorname{Re}\sigma E_{0}^{2}}{2\chi }\frac{L\delta L}{4}\frac{\tanh \left( L/4{{L}_{T}} \right)}{L/4{{L}_{T}}}.
\end{equation}
If the energy sink to phonons is due to deformation potential interaction with acoustic modes, then $L_T\approx \left( v_F/s_a \right){{L}_{fp}}\approx 100{{L}_{fp}}$, where $s_a$ is the sound velocity, ${{v}_{F}}$ is the Fermi velocity, and ${{L}_{fp}}$ is the electron-phonon mean free path. The quantity ${{L}_{T}}$ is then above the length of our device, and the “thermal sink” term can be neglected:
\begin{equation}
\label{Eq-deltaT}
{{T}_{S}}-{{T}_{D}}\approx \frac{\operatorname{Re}\sigma E_{0}^{2}}{2\chi }\frac{L\delta L}{4}.
\end{equation}

To complete the picture, we need to estimate the electric field that leads to the heating of electron system. Electric field in graphene is related to voltage $V_{gr}$ developing on the source-gate section of graphene sheet via $E_0 =2{V_{gr}}/L$. Substituting eq. (\ref{Eq-deltaT}) into (\ref{Eq-UPTE}), we find:
\begin{equation}
\Delta {U_{\rm PTE}}=\left( {{S}_{cont}}-{{S}_{ch}} \right)\frac{\operatorname{Re}\sigma V_{gr}^{2}}{2\chi }\frac{\delta L}{L}.
\end{equation}
If the mechanisms impeding the thermal transport and heat transport are the same, then $\chi$ and $\sigma$ are related via the Wiedemann-Frantz law $\chi /\operatorname{Re}\sigma =\frac{{{\pi }^{2}}}{3}{{\left( \frac{{{k}_{B}}}{e} \right)}^{2}}T$.

The relation between squared voltage across graphene sample and impinging power is linear $V_{gr}^{2}=|{Z}_{\rm eff}| P$, where we have introduced the effective resistance $Z_{\rm eff}$ described in the next section. These estimates lead us to the responsivity ${{R}_{\rm PTE}}=\Delta U_{\rm PTE}/P$ expressed as
\begin{equation}
{{R}_{\rm PTE}}\approx \frac{3}{2{{\pi }^{2}}}\left[ \frac{e}{{k_B}}\left( {{S}_{cont}}-{{S}_{ch}} \right) \right]\frac{e{|Z_{\rm eff}|}}{{k_B}T}\frac{\delta L}{L}.
\end{equation}

\section{Coupling between antenna and graphene channel}

\addDS{In the previous section, we have introduced the effective impedance $Z_{\rm eff}$ relating the power of incoming radiation with the emerging voltage drop across the graphene channel $V_{gr}$. The value of  $Z_{\rm eff}$ depends on the matching between antenna and sample and can be  obtained from the equivalent RF circuit of our device shown in Fig.~\ref{Fig-EqiovSupp} A.}
 \begin{figure}[ht!]
\centering\includegraphics[width=0.8\linewidth]{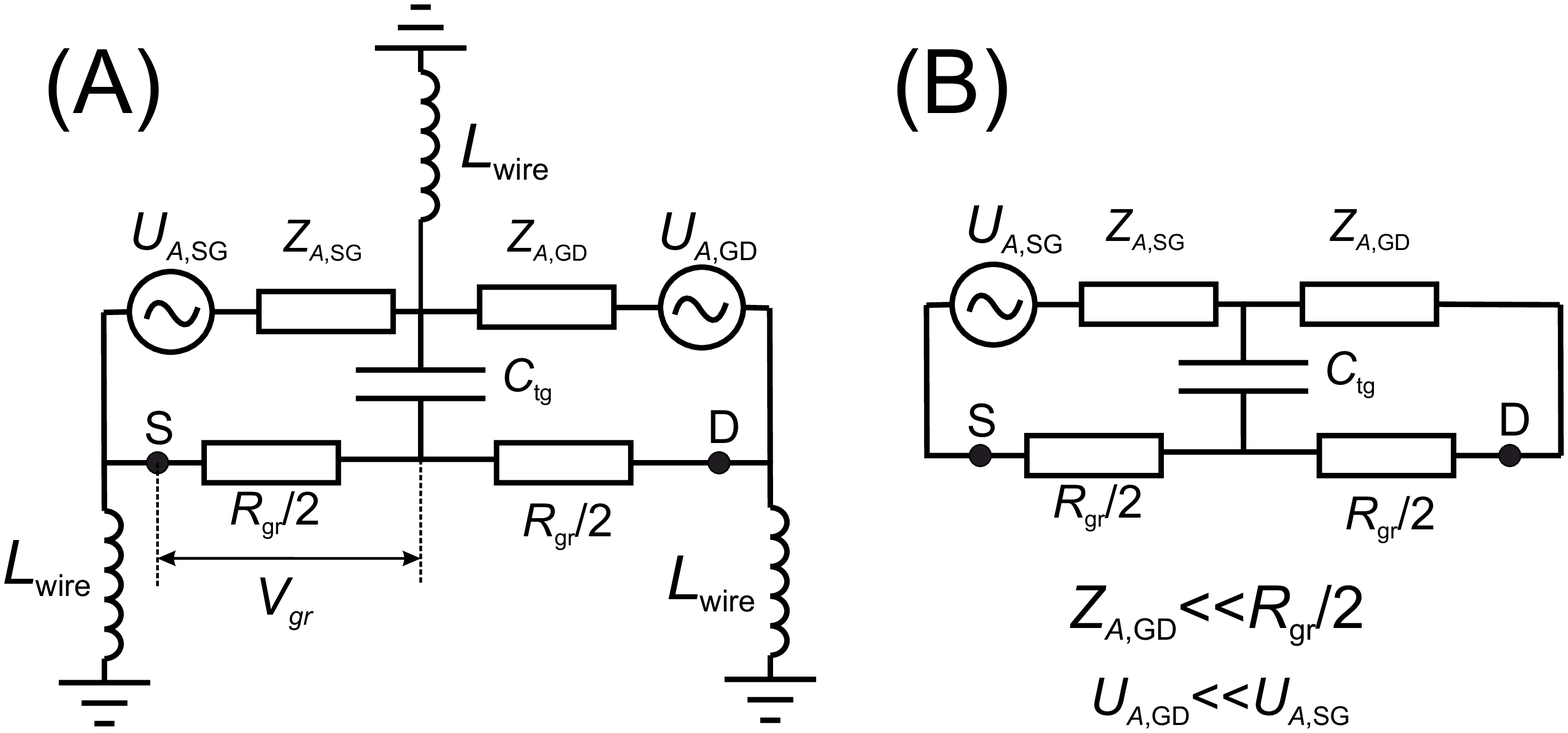}
\label{Fig-EqiovSupp}
\caption{\addDS{(A) Equivalent AC circuit of graphene FET-based detector. Each pair of antenna sleeves acts as a voltage source with open-circuit voltages $U_{A,SG}$ (source-gate) and $U_{A,GD}$ (gate-drain). The impedances between these sleeves are $Z_{A,SG}$ and $Z_{A,GD}$. $L_{wire}$ stands for inductance of contact wires, $C_{tg}$ is the top gate capacitance, $R_{gr}$ is the resistance of graphene channel, $V_{gr}$ is the voltage developed across the left half of graphene sample (B) Simplified equivalent circuit after exclusion of high-inductance wires and voltage source between gate and drain sleeves}}
\end{figure}

\addDS{In this equivalent circuit, each pair of antenna sleeves is modeled as a voltage source. with open-circuit voltages $U_{A,SG}$ (source-gate) and $U_{A,GD}$ (gate-drain). The impedances between these sleeves are $Z_{A,SG}$ and $Z_{A,GD}$.}

\addDS{The inductors ${{L}_{wire}}$ model the connecting wires from antenna sleeves to the measuring circuit. We note that reactive resistance of these wires ${{Z}_{L}}=i\omega {{L}_{\text{wire}}}$ grows with frequency and at $\omega /2\pi =130$ GHz it exceeds all other characteristic resistances in the circuit. Indeed, already for a 1 cm length wire of 100 $\mu$m diameter $L=10$ nH, which corresponds to $Z_L = 8.1 i$ k$\Omega$. Therefore, the high-frequency current does not escape the FET, while the inductances $L_{wire}$ can be safely removed from equivalent circuit. }

\addDS{The origin of asymmetry in the device under study is asymmetric design of the antenna. As the distance between gate and drain sleeves is much smaller than that between gate and source sleeves, the open-circuit voltages in Fig.~\ref{Fig-EqiovSupp} satisfy $U_{A,GD} \ll U_{A,SG}$. As a result, the equivalent circuit simplifies to that shown in Fig.~\ref{Fig-EqiovSupp} B.}

\addDS{It is remarkable that the ratio of currents flowing in the left ($I_l$) and right ($I_r$) sections of the device is small, independent of the impedance between drain and gate sleeves $Z_{A,GD}$. This ratio can be estimated as}
\begin{equation}
\left| \frac{I_l}{I_r} \right|=\left| 1 + \frac{R_{gr}/2 + Z_{A,GD}}{Z_{tg}} \right| ,
\end{equation}
\addDS{here $Z_{tg} = (i\omega C_{tg})^{-1}\approx -1.35 i$ k$\Omega$ is the impedance of the top gate-to-channel capacitor. The impedance $Z_{A,GD}$ is order of free space impedance and can be neglected. The resistance of graphene $R_{gr}/2 = (2...5)$ k$\Omega$, therefore ratio of currents equals $2..4$. This justifies the assumption that the high-frequency current flows predominantly between source and gate terminals leading to highly asymmetric heating (note that Joule power is proportional to the square of current).}

\addDS{ Finally, we relate the voltage across the left part of graphene sample to the antenna voltage}
\begin{equation}
\left| V_{gr} \right|^2 \approx | U_{A,SG}|^{2} \left| \frac{ R_{gr}/2}{ R_{gr}/2+Z_{A,SG}+ Z_{tg} \parallel R_{gr}/2} \right|^2,
\end{equation}
\addDS{ where the $\parallel$ sign stands for impedance of elements in parallel connection. The open-circuit antenna voltage $U_{A,SG}$ is related to the electric field strength in impinging wave ${{E}_{0}}$ via an effective antenna height $U_{A,SG} = {h_{\rm eff}}{E_0}$. On the other hand, the electric field is related to the intensity $I$ via free-space impedance, ${Z_0}$ as  $E_{0}^{2}=2{Z_0}I/n_{\rm Si}=2{Z_0}P/A$ , where $A$ is the geometrical area of the antenna, and $n_{\rm Si}\approx 3.5$ is the refractive index of the substrate. Combining these expressions, we find that the  ac voltage across the sample is related to impinging power via}
\begin{equation}
{\left| V_{gr} \right|^2}\approx |U_{A,SG}|^2 
\frac{h_{\rm eff}^2}{A}
\frac{2Z_0P/n_{\rm Si}}{\left| 1+\frac{Z_{A,SG} +Z_{tg} \parallel R_{gr}/2 }{R_{gr}/2} \right|^2}.
\end{equation}
\addDS{ Hence, the effective impedance}
\begin{equation}
Z_{\rm eff} = \frac{2 Z_0}{n_{\rm Si}}
\frac{h_{\rm eff}^2/A}
{{
\left| 1+\frac{Z_{A,SG} + Z_{tg} \parallel R_{gr}/2 }{R_{gr}/2} 
\right|}^2}.
\end{equation}
\addDS{ The denominator of this expression is responsible for the reduction of the voltage at non-optimal matching. We note that the voltage developed across the device increases up to saturation with increasing the sample resistance ${{R}_{gr}}$, while the absorbed power ${{\left| V_{gr} \right|}^{2}}/{{R}_{gr}}$ naturally has an optimum under matching conditions. The radiative resistance of a spiral antenna is~\cite{milligan2005modern} }
\begin{equation}
Z_{A,SG} \approx \frac{Z_0}{2}\sqrt{\frac{2}{1+n^2_{\rm Si}}} \approx 74 \,\Omega.
\end{equation}
\addDS{ Also, typically $h_{\rm eff}^{2} \sim A$. Using the experimentally obtained value of graphene resistance away from CNP $R_{gr}/2 \approx 2$ k$\Omega$, we estimate the effective impedance as 140 $\Omega$. This agrees well with the value $Z_{\rm eff} \approx 100$ $\Omega$ obtained from comparison of experimentally measured responsivity with the PTE model in the main text.}

\section{Sign of the photovoltage}
In this section we show that both PTE and DS rectification scenario yield the same sign of the photovoltage. For simplicity, we discuss the case of the n-doped graphene channel, the extension for the p-doped case can be obtained conversely to what is discussed below. 

In the model of resistive self-mixing (DS overdamped photoresponse) the sign of the photovoltage can be understood as following. In the first half-period of the gate voltage oscillation, i.e. when the voltage is positive with respect to the channel, electrons enter the channel from the source. In the second half-period, electrons leave the channel. Since the carrier density defines the conductivity, the latter is larger in the first half of the period. This means that the rectified electron flux is in the direction from the source to the drain terminals. Since the drain is opened, the DC drain voltage developing in the steady state must be negative to stop this flux. 

The sign of the PTE is more intuitive. Heating of the n-doped sample from the source side leads to electron diffusion from hot to cold end, i.e. from the source to drain. Accumulation of the negative charge carriers at the drain leads to the negative drain potential.

To conclude, both PTE and DS rectification mechanisms result in the same sign of the photovoltage.

\end{widetext}

\end{document}